\pdfoutput=1
\documentclass[12pt]{article}
\usepackage{times}
\usepackage{graphicx}
\usepackage{amsmath,amssymb}
\usepackage{textcomp}
\usepackage[margin=1in]{geometry}
\usepackage[small,bf]{caption}
\usepackage{longtable}
\usepackage[square,numbers,sort]{natbib}
\usepackage[colorlinks,linkcolor=black,bookmarksopen,bookmarksnumbered,
citecolor=black,urlcolor=black]{hyperref}

\graphicspath{{figs/wct/examples/},{figs/wct/meshes/},{figs/wct/misc/}}

\newcommand{\RR}{\ensuremath{\mathbb{R}}}

\newlength{\dotlen}

\title{A Dihedral Acute Triangulation of the Cube}

\author{Evan VanderZee
  \thanks{Department of Mathematics, University of Illinois at
    Urbana-Champaign (vanderze@illinois.edu)}
  \and Anil N. Hirani
  \thanks{Department of Computer Science,
    University of Illinois at Urbana-Champaign
    (hirani@cs.uiuc.edu)}
  \and Vadim Zharnitsky
  \thanks{Department of Mathematics, University of Illinois at
    Urbana-Champaign (vz@math.uiuc.edu)}
  \and Damrong Guoy
  \thanks{Synopsys Inc. (Damrong.Guoy@synopsys.com)}
}
\date{}

\begin{document}
\maketitle

\vspace{-1em} 
\abstract{It is shown that there exists a dihedral acute triangulation
  of the three-dimensional cube.  The method of constructing the acute
  triangulation is described, and symmetries of the triangulation are
  discussed.}
%\keywords{dihedral acute -- acute triangulation -- cube}

\section{Introduction}
\label{sec:intro}

Interest in acute triangulation of polyhedra dates back to the 1960s
at least; when geometers were first working on proving that abstract
polyhedra could be realized geometrically, acute triangulations of
polyhedra played a role in the solution~\cite{BuZa1960}. In the 1970s,
Ciarlet and Raviart showed that a finite element solution of a
reaction-diffusion problem satisfies a discrete maximum principle if
the triangulation is acute and satisfies some other geometric
conditions~\cite{CiRa1973}. Keen interest in acute and nonobtuse
triangulations continues today, as evidenced by a recent review
article on the subject~\cite{BrKoKrSo2009}.

The review article poses the specific problem of obtaining dihedral
acute triangulations of domains in high-dimen\-sional spaces. The
problem also appears in unpublished lecture notes of
Pak~\cite{Pak2009}. Pak mentions that in $\RR^{5}$ and in
higher-dimen\-sional spaces, there is no acute triangulation of the
hypercube, leaving the proof to the reader. The problem is a
combinatorial one, and a proof is given in the literature by
K{\v{r}}{\'\i}{\v{z}}ek~\cite{Krizek2006}. The acute triangulation of
$\RR^{3}$ and of infinite slabs in $\RR^{3}$ was solved by Eppstein,
Sullivan, and {\"U}ng{\"o}r~\cite{EpSuUn2004}, who also stated that
acute triangulation of the cube is an open problem. Brandts, Korotov,
and K{\v{r}}{\'\i}{\v{z}}ek mention the problem in a recent paper that
improves the results of Ciarlet and Raviart but retains the
requirement that the finite element mesh be an acute
triangulation~\cite{BrKoKr2008}. Saraf also indicates that acute
triangulation of the cube is an open problem~\cite{Saraf2009}.

Acute triangulation can be thought of as an improvement of a nonobtuse
triangulation. However, in general acute triangulation is a much more
challenging problem than nonobtuse triangulation. As noted above,
there is no acute triangulation of the hypercube in $\RR^{5}$ or in
higher-dimen\-sional spaces, but there is a nonobtuse triangulation of
the hypercube in any dimension~\cite{BeChEpRu1995}. The construction
uses path simplices, and the basic idea is to add a main diagonal of
the hypercube. This construction works in $\RR^{3}$, as well,
producing a nonobtuse triangulation of the cube with six congruent
nonobtuse tetrahedra fitting together around a main diagonal. A
nonobtuse triangulation with five tetrahedra is also possible;
removing the regular tetrahedron whose vertices are four pairwise
nonadjacent corners of the cube leaves four nonobtuse tetrahedra. In
contrast, some simple computations with Euler's formula show that any
acute triangulation of the cube must have more than $100$ tetrahedra.

This paper shows that the cube in $\RR^3$ does have an acute
triangulation. In fact, it has infinitely many acute triangulations.
The acute triangulation described in this paper has $1370$ tetrahedra.
Details about it are given in Sec.~\ref{sec:acutetrngltn}, along with
some statistics that show the superior quality of the triangulation.
The maximum dihedral angle is around $84.65$\textdegree, well within
the range of acute, and the minimum dihedral angle is a nice
$35.89$\textdegree. Section~\ref{sec:construct} describes the
computer-assisted construction of this acute triangulation of the
cube; a hand construction was combined with mesh optimization to build
the mesh. The triangulation has some symmetries, which are discussed
in Sec~\ref{sec:symmetry}. The symmetries greatly reduce the number of
distinct tetrahedra from $1370$ to $82$, and can be used to generate
the full set of $277$ vertices from just $26$ of them.

\section{The Acute Triangulation}
\label{sec:acutetrngltn}

We present the first-known acute triangulation of the cube as a
triangulation of a cube centered at the origin with corner vertices at
$(\pm 1, \pm 1, \pm 1)$. The coordinates of the vertices of the
triangulation are listed in Appendix~\ref{sec:vtxcoord}. They are also
available online, along with code to compute the
angles~\cite{VaHiZhGu2009a}. The mesh connectivity is given by the
Delaunay triangulation of the set of vertices. It has been shown that
an acute triangulation in three dimensions is not necessarily a
Delaunay triangulation~\cite{EpSuUn2004}. This acute triangulation of
the cube, however, is not only Delaunay, but also is one for which
each tetrahedron properly contains its circumcenter\footnote{This
  property does not hold in general for acute triangulations. For a
  tetrahedral mesh in $\RR^{3}$, if each tetrahedron contains its
  circumcenter, the mesh must be
  Delaunay~\cite{Rajan1994}~\cite{ScSp1999}~\cite{VaHiGuRa2008}, so
  any non-Delaunay acute triangulation does not have this property.},
i.e., the triangulation is $3$-well-centered~\cite{VaHiGuRa2008}.

\begin{figure}
  \centering
  \includegraphics[width=180pt, trim=485pt 195pt 413pt 142pt, clip]%
  {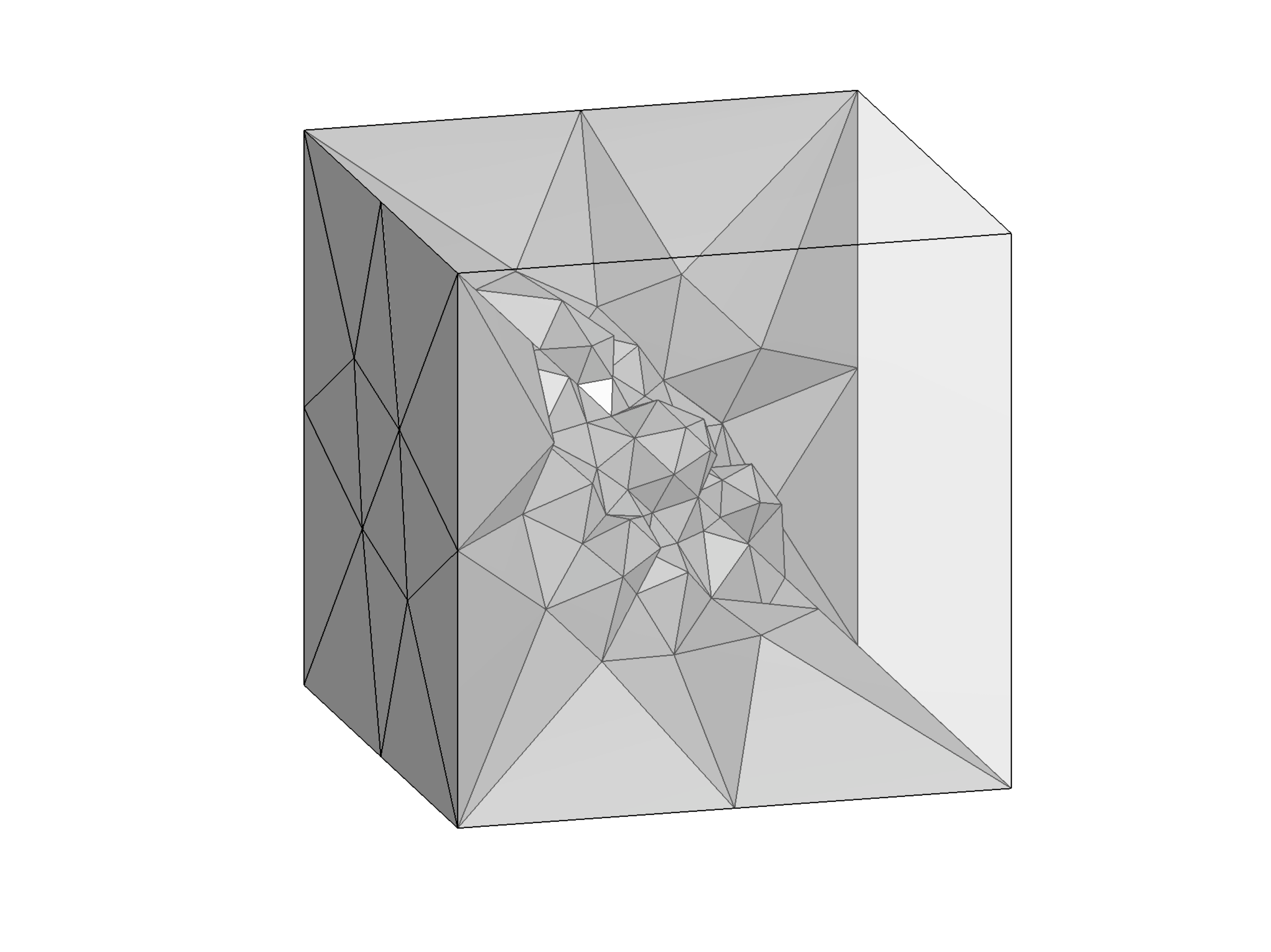}%
  \hspace{40pt}%
  \includegraphics[width=194pt, trim=435pt 192pt 365pt 139pt, clip]%
  {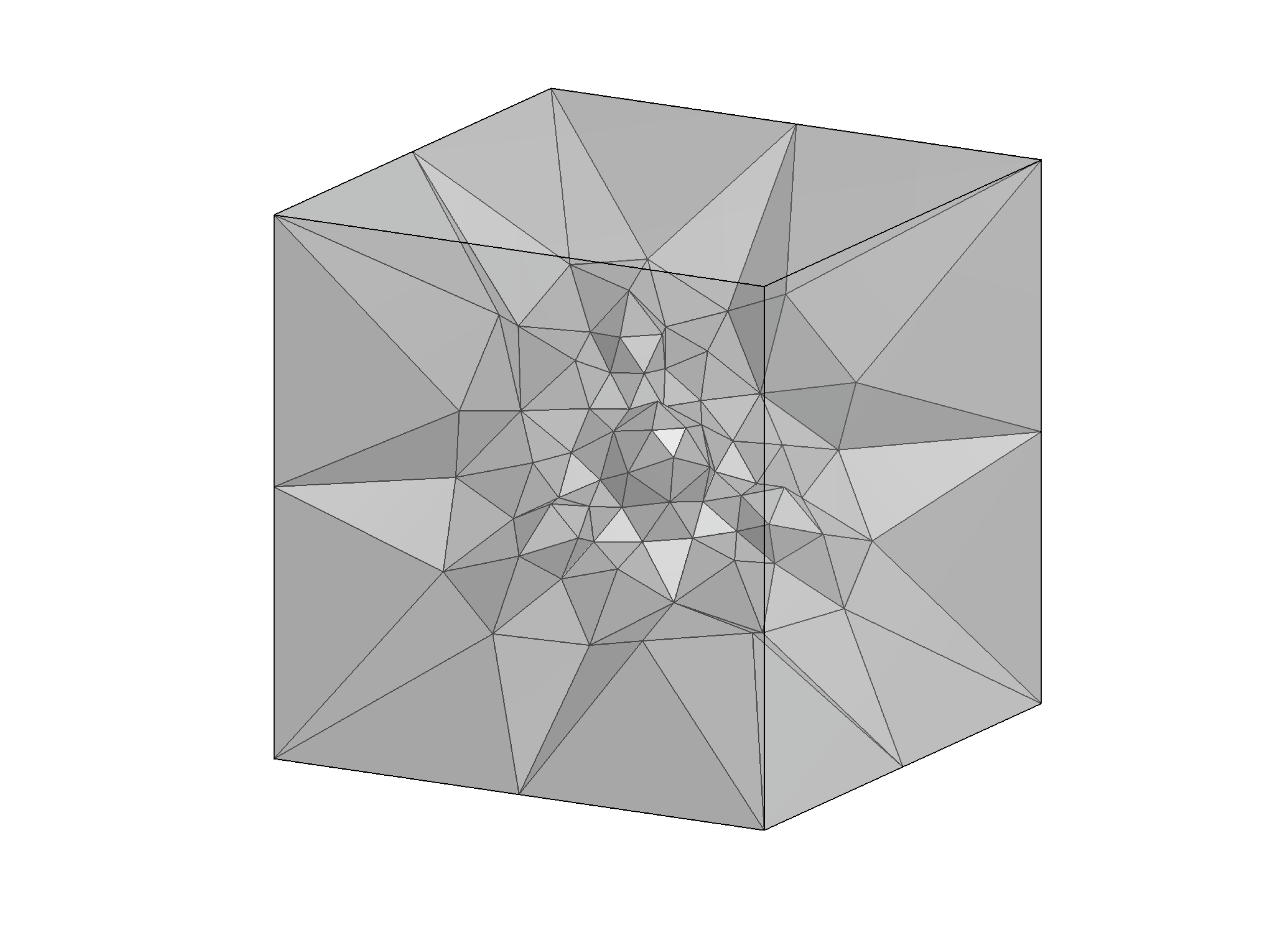}%
  \caption{Two views of a cutaway section of the first-known acute
    triangulation of the cube.  The view at right is a
    $45$\textdegree\ rotation about the $z$-axis from the view at
    left.  On the left a $14$-triangle triangulation of one of the
    square faces of the cube is visible.  This $14$-triangle
    triangulation of the square is used on each face of the cube.}
\label{fig:cubecutaway}
\end{figure}

\begin{figure}
%
% Figures in a row
%
\begin{minipage}[b]{140pt}
  \centering
  \includegraphics[width=140pt, trim=198pt 296pt 174pt 320pt, clip]%
  {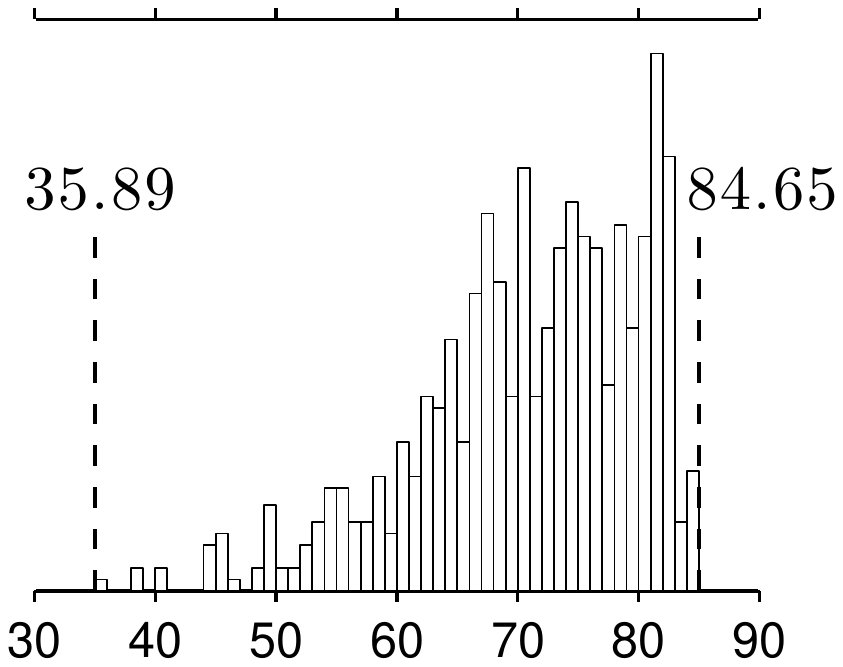}%
\end{minipage}%
\hspace{10pt}
\begin{minipage}[b]{140pt}
  \centering
  \includegraphics[width=140pt, trim=198pt 297pt 185pt 318pt, clip]%
  {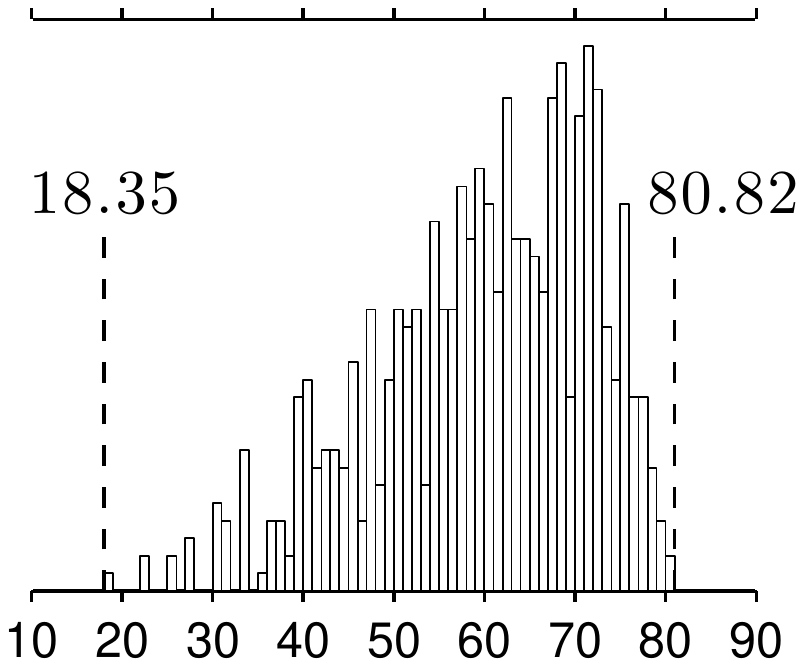}%
\end{minipage}%
\hspace{10pt}
\begin{minipage}[b]{140pt}
  \centering
  \includegraphics[width=140pt, trim=199pt 297pt 194pt 317pt, clip]%
  {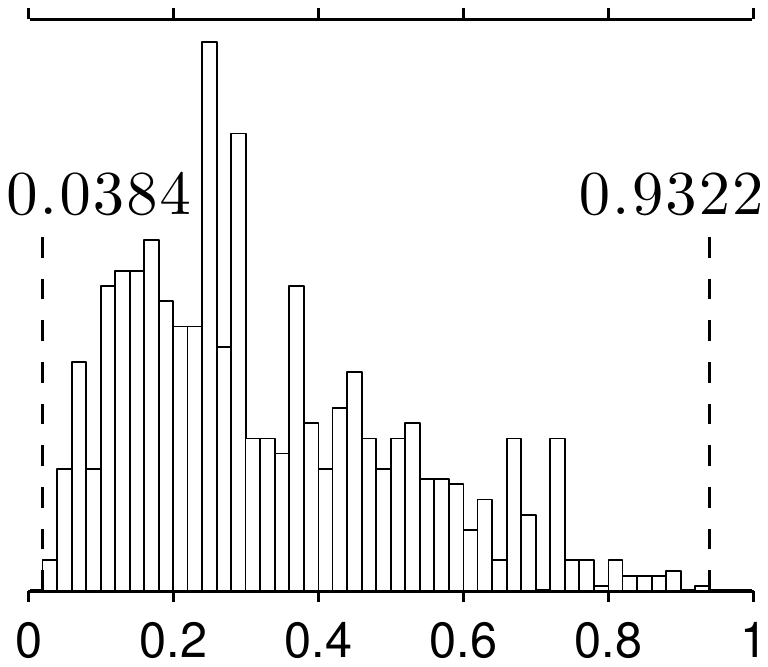}%
\end{minipage}\\[-4pt]
% 
% Captions below
% 
\begin{minipage}[t]{140pt}
  \caption{A histogram of the dihedral angles of the acute
    triangulation of the cube.}
  \label{fig:histdihedral}
\end{minipage}%
\hspace{10pt}
\begin{minipage}[t]{140pt}
  \caption{A histogram of the face angles of the acute triangulation
    of the cube.}
  \label{fig:hist2wc}
\end{minipage}%
\hspace{10pt}
\begin{minipage}[t]{140pt}
  \caption{A histogram of the tetrahedron $h/R$ values of the acute
    triangulation of the cube.}
  \label{fig:hist3wc}
\end{minipage}%
\end{figure}

Figure~\ref{fig:cubecutaway}, a cutaway view of the acute
triangulation of the cube, visually shows the high quality of the
triangulation. Figures~\ref{fig:histdihedral}
through~\ref{fig:hist3wc} give quantitative evidence of the quality of
the triangulation. Each figure is a histogram of some quantitative
measurement of the quality of tetrahedra. In each case, the histogram
summarizes all of the values of the quantity in the mesh. For
instance, the histogram of dihedral angles shows all of the dihedral
angles, not just the maximum dihedral angle of each tetrahedron. The
$h/R$ values summarized in Fig.~\ref{fig:hist3wc}, which may be less
familiar to readers than the other measurements, are related to
$3$-well-centeredness. The range of the quantity $h/R$ over all
tetrahedra is $(-1, 1)$, with $3$-well-centered tetrahedra having all
values in the range $(0, 1)$. The $h/R$ values in a regular
tetrahedron are all $1/3$. See~\cite{VaHiGuRa2008}
or~\cite{VaHiGu2008} for more details.

Combinatorics plays an important role in acute triangulation, so we
briefly mention some of the combinatorial statistics of the acute
triangulation of the cube. There are $277$ vertices, $1688$ edges, and
$1370$ tetrahedra. Of the edges, $126$ are boundary edges. Of the
interior edges, $1506$ have the minimum possible number of incident
tetrahedra for an acute triangulation, i.e., $5$ tetrahedra, and the
remaining $56$ each have $6$ incident tetrahedra. For the vertices,
$44$ are on the boundary, and $233$ are interior. A large majority of
the interior vertices ($200$ of them) have icosahedral neighborhoods,
thus they have $12$ incident edges. There are $10$ vertices with $14$
incident edges, $18$ vertices with $15$ incident edges, $4$ vertices
with $16$ incident edges, and $1$ vertex---the central vertex located
at $(0, 0, 0)$---with $22$ incident edges.

The high degree central vertex can be replaced with a regular
tetrahedron to give a combinatorially different acute triangulation of
the cube, one with $1387$ tetrahedra. To obtain this triangulation,
replace the vertex at the origin with the four vertices at $(-0.05,
-0.05, 0.05)$, $(-0.05, 0.05, -0.05)$, $(0.05, -0.05, -0.05)$, and
$(0.05, 0.05, 0.05)$ and compute the Delaunay triangulation of the new
vertex set. The result is acute and completely well-centered. We see
that there are at least two combinatorially distinct acute
triangulations of the cube conforming to the same surface
triangulation.

\section{Method of Construction}
\label{sec:construct}

The basic methodology for the construction was one of an advancing
front. It is absolutely necessary to have an acute surface
triangulation, since an acute tetrahedron has acute
facets~\cite{EpSuUn2004}, and more generally, all facets of an acute
simplex are acute~\cite{BrKoKr2007}. We began with a high-quality
acute surface triangulation of the cube; the midpoint of each edge of
the cube was added, and each face was triangulated with a
$14$-triangle acute triangulation that conforms to this boundary and
has a maximum face angle around $73.3$\textdegree. On the left side of
Fig.~\ref{fig:cubecutaway} one can see a $14$-triangle acute
triangulation of the square on one of the faces of the cube.

Starting from the acute surface triangulation, we built inward,
carefully adding vertices and tetrahedra to satisfy the combinatorial
constraints. That is, each edge of the triangulation coinciding with
an edge of the cube must have at least two incident tetrahedra, each
edge of the triangulation lying in a facet of the cube must have at
least three incident tetrahedra, and each interior edge must have at
least five incident tetrahedra. The addition of vertices and tetrahedra
was performed by hand with the frequent computation of the Delaunay
triangulation to help get the proper mesh connectivity.

After each layer or partial layer was constructed by hand, the mesh
was optimized to obtain a set of acute tetrahedra conforming to the
boundary of the cube. The optimization did not explicitly seek a
dihedral acute triangulation, but instead tried to make the meshes
completely well-centered. (This type of optimization was introduced
for two dimensions in~\cite{VaHiGuRa2007} and later generalized to
higher dimensions in~\cite{VaHiGuRa2008}.) At each layer, a moderately
aggressive version of the optimization yielded a mesh that was both
completely well-centered and dihedral acute. In most cases, more
aggressive optimization produced a mesh that was well-centered but not
acute, and less aggressive optimization produced a mesh that was
neither well-centered nor acute. When a dihedral acute and completely
well-centered mesh was obtained from the optimization, a new layer
consisting of more tetrahedra and vertices was added by hand.

Eventually this process reached a stage in which all of the edges on
the internal boundary already had three incident tetrahedra. This and
the rest of the combinatorics and geometry worked out so that adding a
vertex at the center of the cube and computing the Delaunay
triangulation produced an acute, completely well-centered
triangulation of the cube.

When the acute triangulation was first obtained and optimized relative
to a cost function based on well-centered\-ness, it had a maximum
dihedral angle around $87.8$\textdegree. Later some additional
optimization was applied that directly optimized the dihedral angle as
well as optimizing for well-centeredness, and the symmetry discussed
in the next section was enforced exactly. The additional optimization
allowed boundary vertices to move constrained to the surfaces of the
cube. This optimization produced the final mesh presented in this
paper, except that the vertices were rounded to the nearest $.001$ for
ease of presentation.

\section{Symmetries of the Triangulation}
\label{sec:symmetry}

The acute triangulation of the cube presented in this paper has an
$S_{4}$ symmetry group. More precisely, it has all of the symmetries
of a regular tetrahedron whose vertices are four pairwise nonadjacent
corners of the cube. Consider, for instance, the regular tetrahedron
with vertices $(-1, -1, -1)$, $(-1, 1, 1)$, $(1, -1, 1)$,
and $(1, 1, -1)$. Each of the $24$ symmetries of this regular
tetrahedron---rotations or reflections in $\RR^{3}$ that map the
tetrahedron to itself---is a symmetry that maps this acute
triangulation of the cube to itself.

In fact, it is possible to use these symmetries to construct the full
set of $277$ vertices from just $26$ of them. There are multiple ways
to do this, one of which is the following. Take the $26$ vertices in
the list in Table~\ref{table:vtxcoordsym1} in
Appendix~\ref{sec:vtxcoord}. These $26$ vertices are the vertices that
lie in the $1/24$th of the cube specified by the inequalities
$y \ge -1$, $x \ge y$, $x \le z$, and $x \le -z$.

We will transform this initial set of vertices using the orthogonal
matrices
\[
A_{1} = \left[
  \begin{matrix}0 & 0 & -1\\ 0 & 1 & 0\\ -1 & 0 & 0\end{matrix}\right]\quad
A_{2} = \left[
  \begin{matrix}0 & 0 & 1\\ 0 & 1 & 0\\ 1 & 0 & 0\end{matrix}\right]\quad
A_{3} = \left[
  \begin{matrix}-1 & 0 & 0\\ 0 & -1 & 0\\ 0 & 0 & 1\end{matrix}\right]\quad
A_{4} = \left[
  \begin{matrix}0 & 1 & 0\\ 0 & 0 & 1\\ 1 & 0 & 0\end{matrix}\right]
  \,\text{.}
\]
Each of these matrices is a symmetry of the aforementioned regular
tetrahedron and a symmetry of the cube. Matrix $A_{1}$ is a reflection
through the plane $x = -z$. Matrix $A_{2}$ is a reflection through the
plane $x = z$. Matrix $A_{3}$ is a $180$\textdegree\ rotation about
the $z$-axis, which could also be thought of as reflection through the
$z$-axis. Finally, $A_{4}$ is a rotation about the main diagonal of
the cube passing through $(-1, -1, -1)$ and $(1, 1, 1)$. Looking along
this diagonal from $(-1, -1, -1)$ towards $(1, 1, 1)$, the rotation is
$120$\textdegree\ counterclockwise.

\begin{figure}
  \centering
  \includegraphics[width = 100pt, trim = 534pt 206pt 462pt 150pt, clip]
  {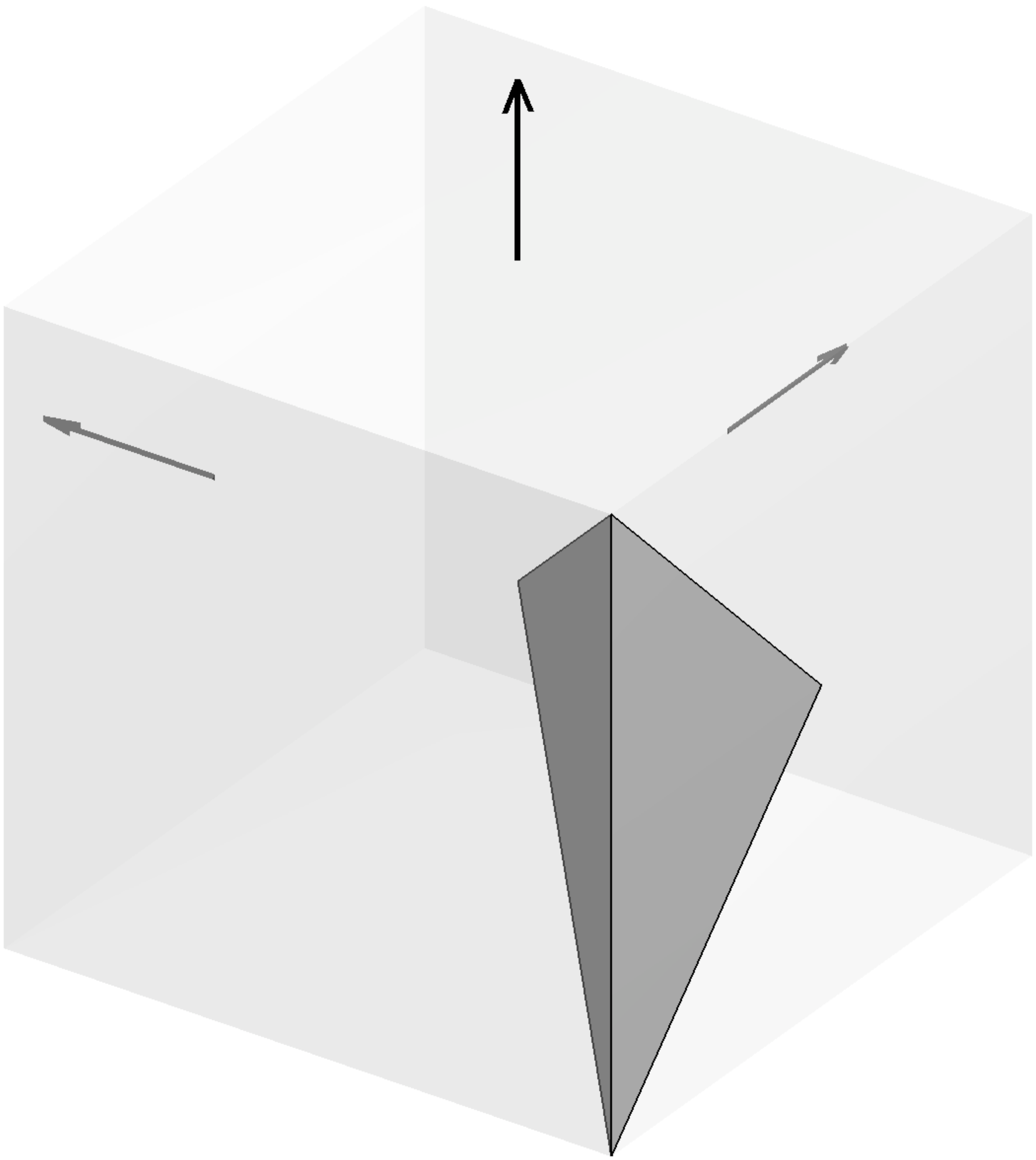}%
  \begin{picture}(0, 0)
    \put(-26, 79){$x$}
    \put(-92, 75){$y$}
    \put(-58, 99){$z$}
  \end{picture}%
  \hspace{50pt}%
  \includegraphics[width = 100pt, trim = 534pt 206pt 462pt 150pt, clip]
  {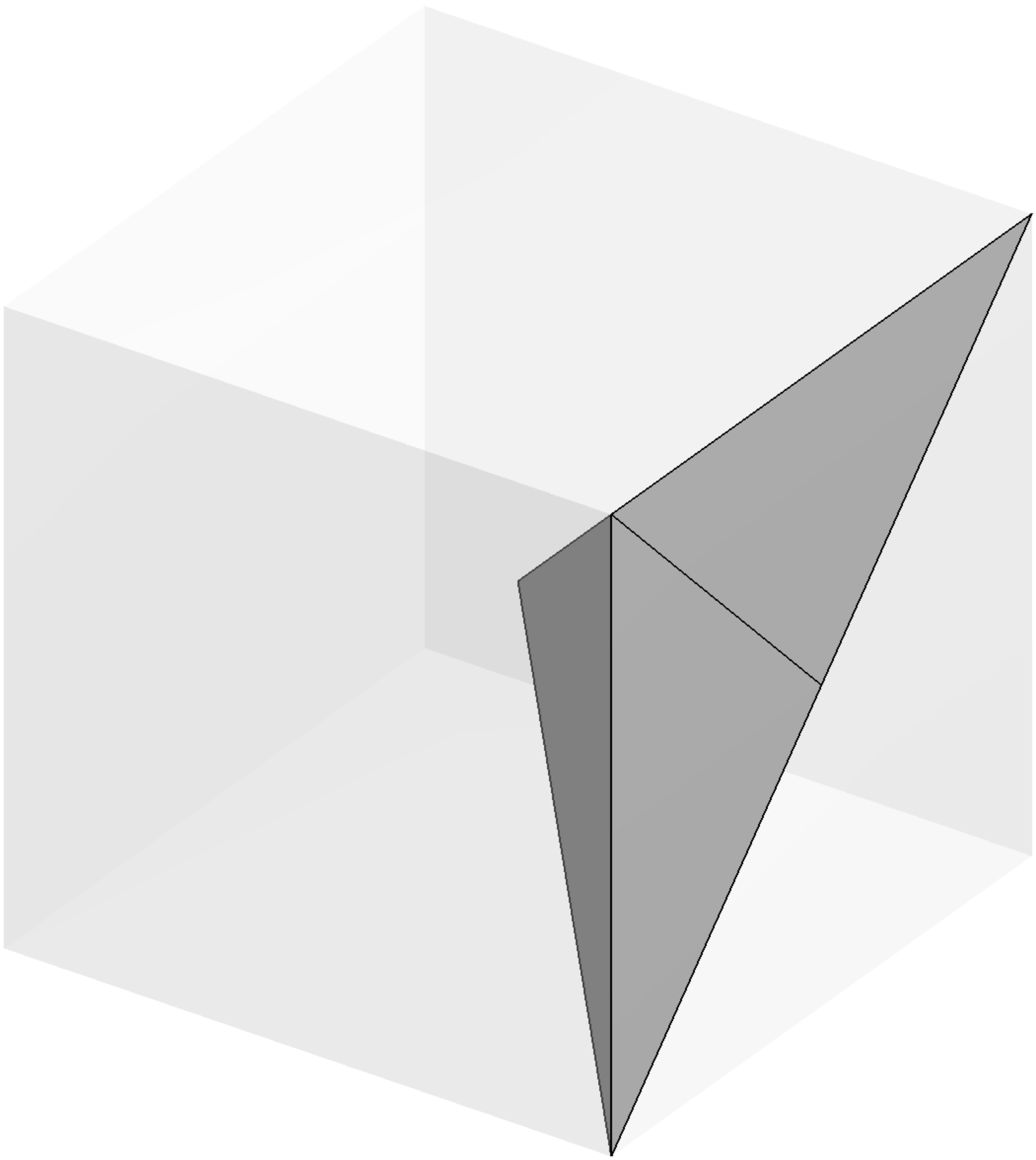}%
  \hspace{50pt}%
  \includegraphics[width = 100pt, trim = 534pt 206pt 462pt 150pt, clip]
  {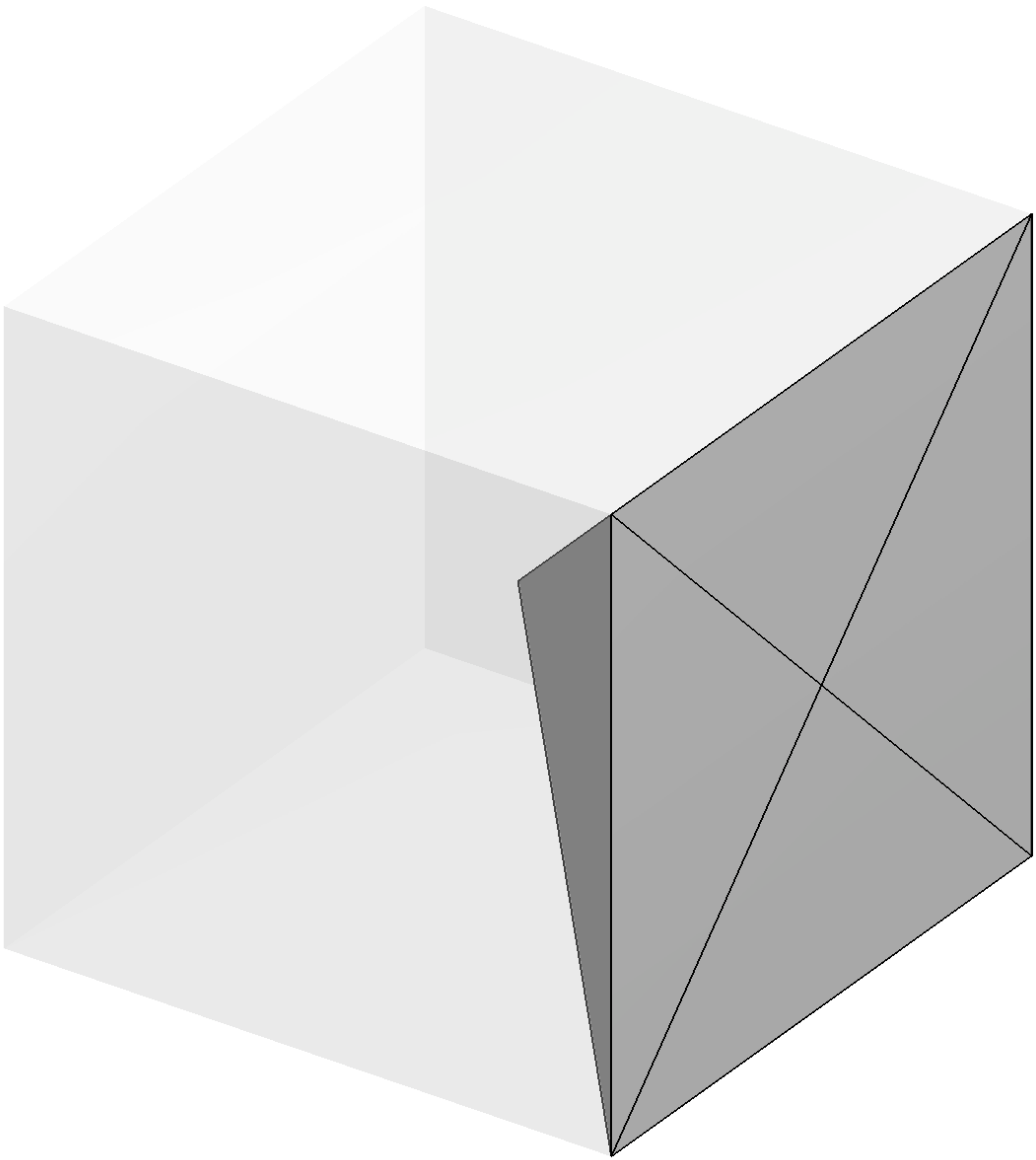}\\[10pt]
  \includegraphics[width = 100pt, trim = 534pt 206pt 462pt 150pt, clip]
  {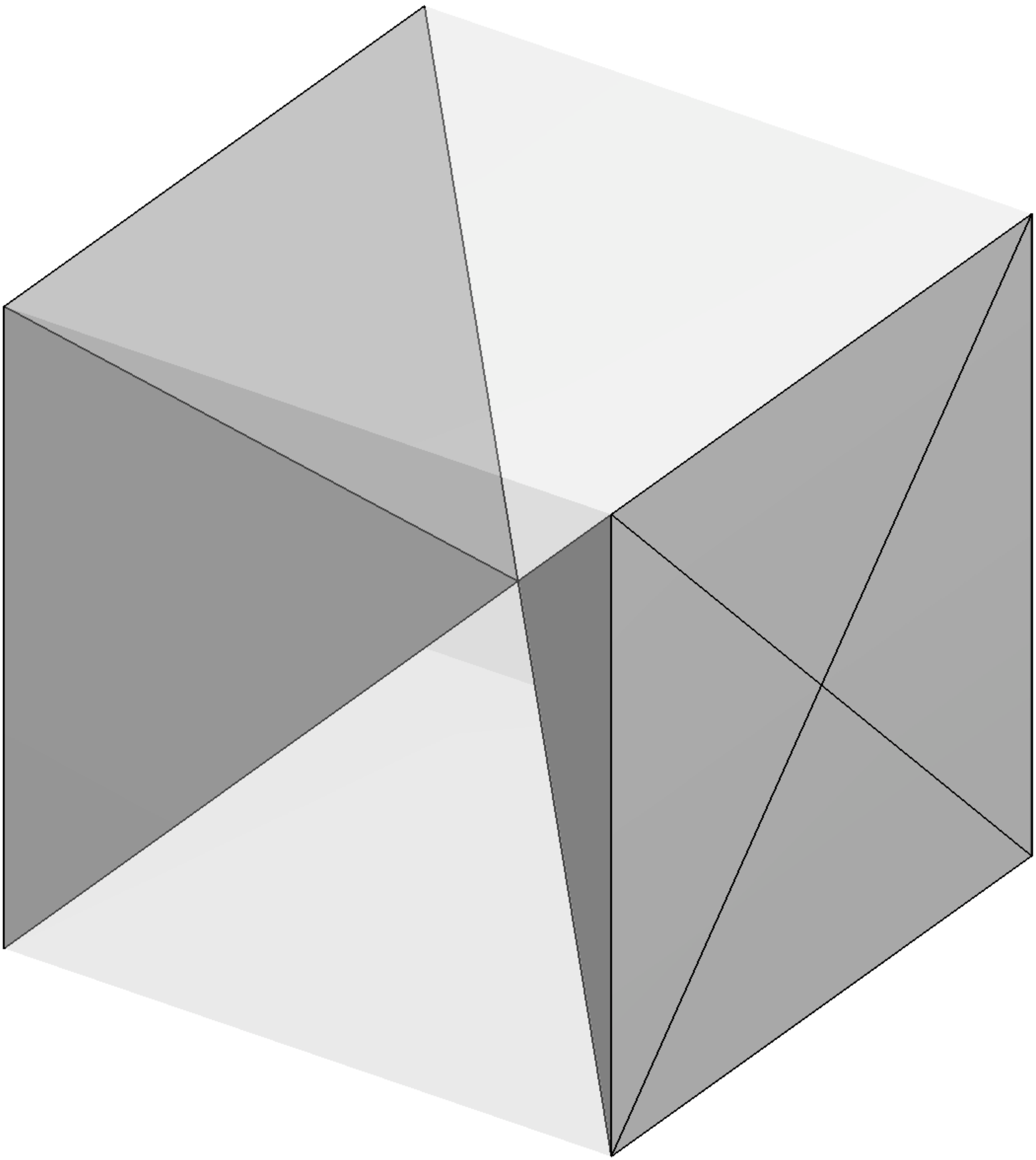}%
  \hspace{50pt}%
  \includegraphics[width = 100pt, trim = 534pt 206pt 462pt 150pt, clip]
  {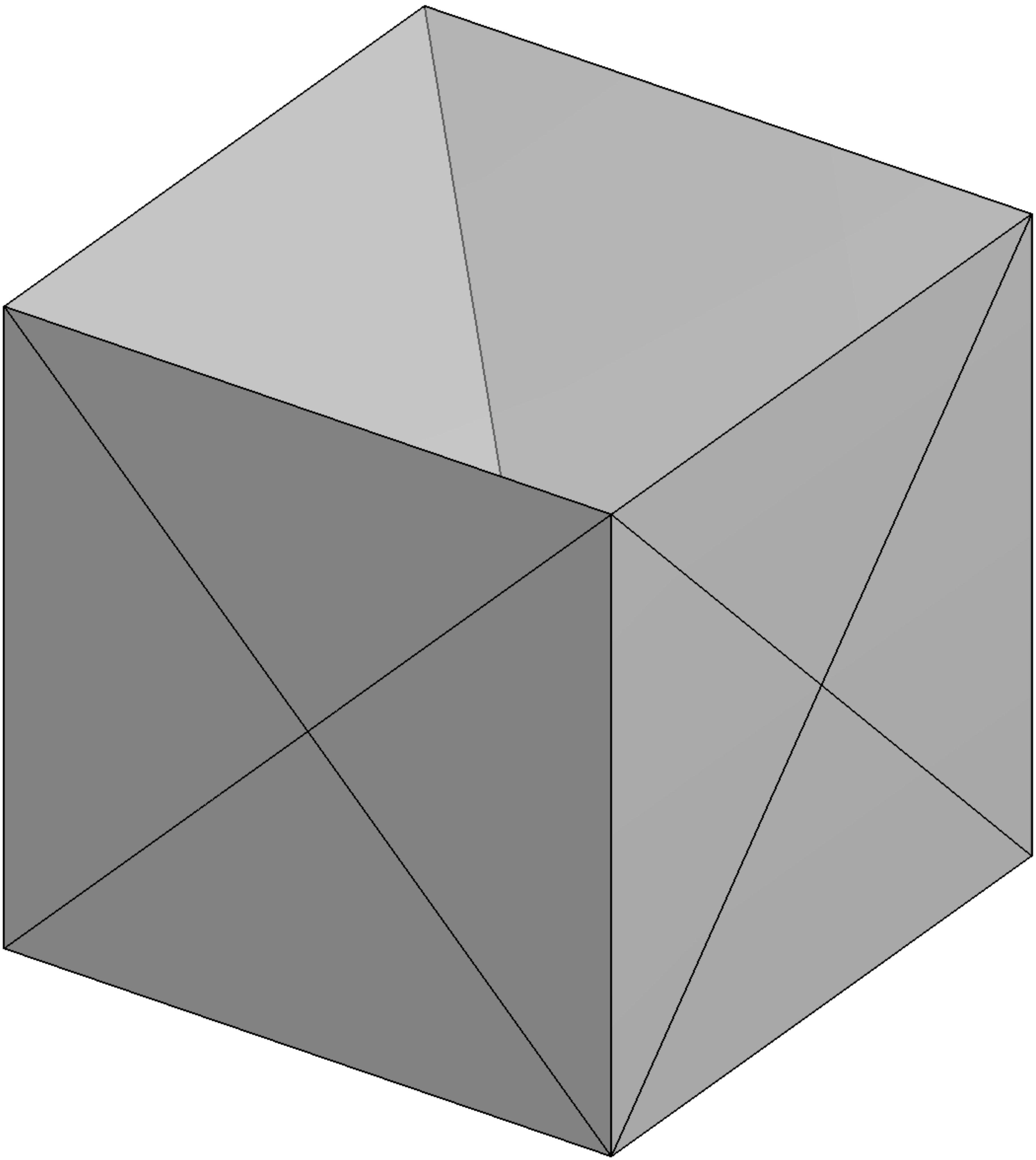}%
  \hspace{50pt}%
  \includegraphics[width = 100pt, trim = 534pt 206pt 462pt 150pt, clip]
  {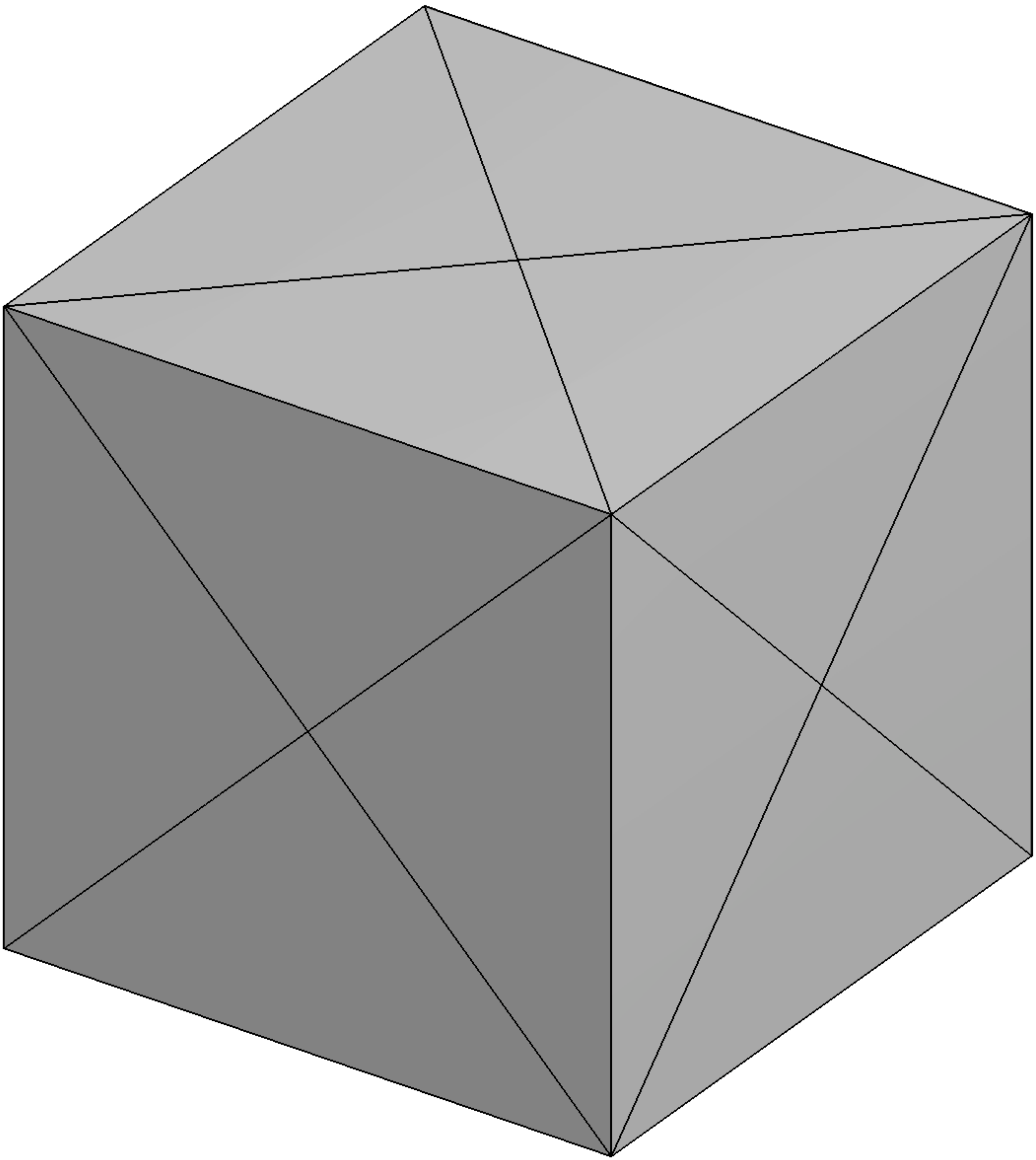}
  \caption{Starting from a set of $26$ vertices lying in $1/24$th of
    the cube, the full set of $277$ vertices can be generated using
    symmetries of the acute triangulation of the cube.  This figure
    shows one way that the symmetries can be used to generate the full
    vertex set.  Starting from the $1/24$th of the cube bounded by the
    planes $y = -1$, $x = y$, $x = z$, and $x = -z$ (top left), one
    can fill the cube by performing the following steps in order.
    Reflect across the plane $x = -z$ (top center).  Reflect across
    the plane $x = z$ (top right).  Rotate by $180$\textdegree\ about
    the $z$-axis (bottom left).  Simultaneously rotate by
    $120$\textdegree\ (bottom center) and $240$\textdegree\ (bottom
    right) about the main diagonal through $(-1, -1, -1)$ and $(1, 1,
    1)$.}
  \label{fig:symmetries}
\end{figure}
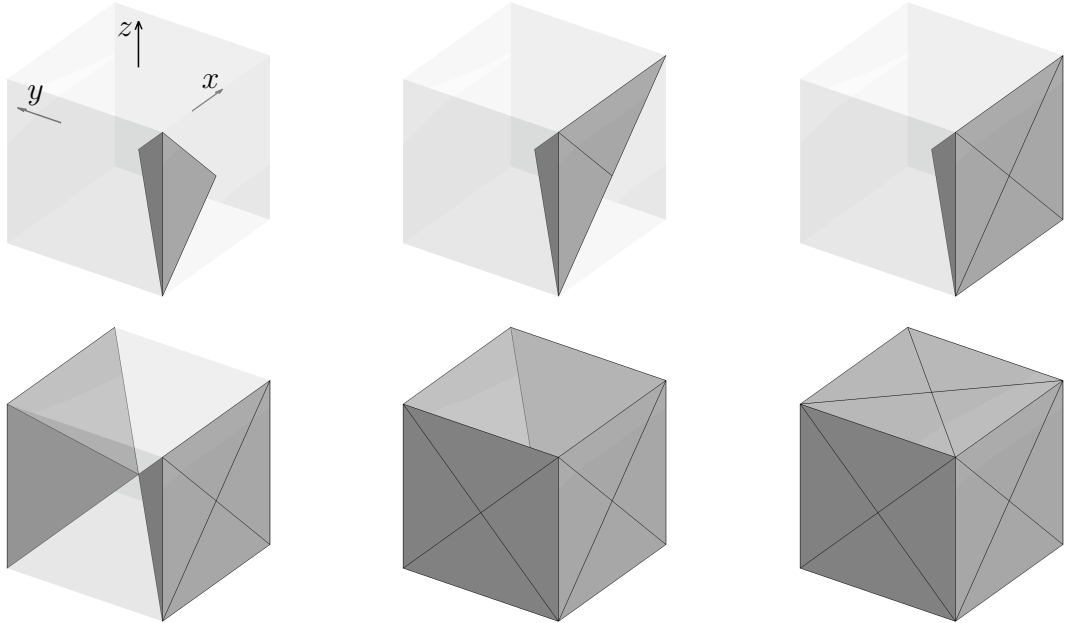

Figure~\ref{fig:symmetries} shows how these symmetries fill the cube
starting from the generating $1/24$th section of the cube. Taking the
initial set of $26$ vertices that lie in the generating section of the
cube, we apply matrix $A_{1}$ to obtain a set of vertices that lie in
$1/12$th of the cube. Then we apply $A_{2}$ to the new vertex set,
obtaining a set of vertices lying in $1/6$th of the cube. The region
containing the vertices is now a pyramid over the face $y = -1$ with
apex at the origin. (Top right in Fig.~\ref{fig:symmetries}.) To this
vertex set we apply $A_{3}$ and cover $1/3$rd of the cube. Finally we
apply both $A_{4}$ and $A_{4}^{2}$---rotating by both
$120$\textdegree\ and $240$\textdegree\ about a main diagonal---to
obtain a set of vertices that covers the full cube. A large number of
vertices in this vertex set are duplicates of each other, but when all
of the duplicates are removed, there are $277$ vertices that remain.
The listing of vertex coordinates in Appendix~\ref{sec:vtxcoord} is
divided into separate tables according to the way these symmetries
generate the full set of vertices.

\begin{figure}
  \centering
  \includegraphics[width=180pt, trim=350pt 164pt 276pt 114pt, clip]%
  {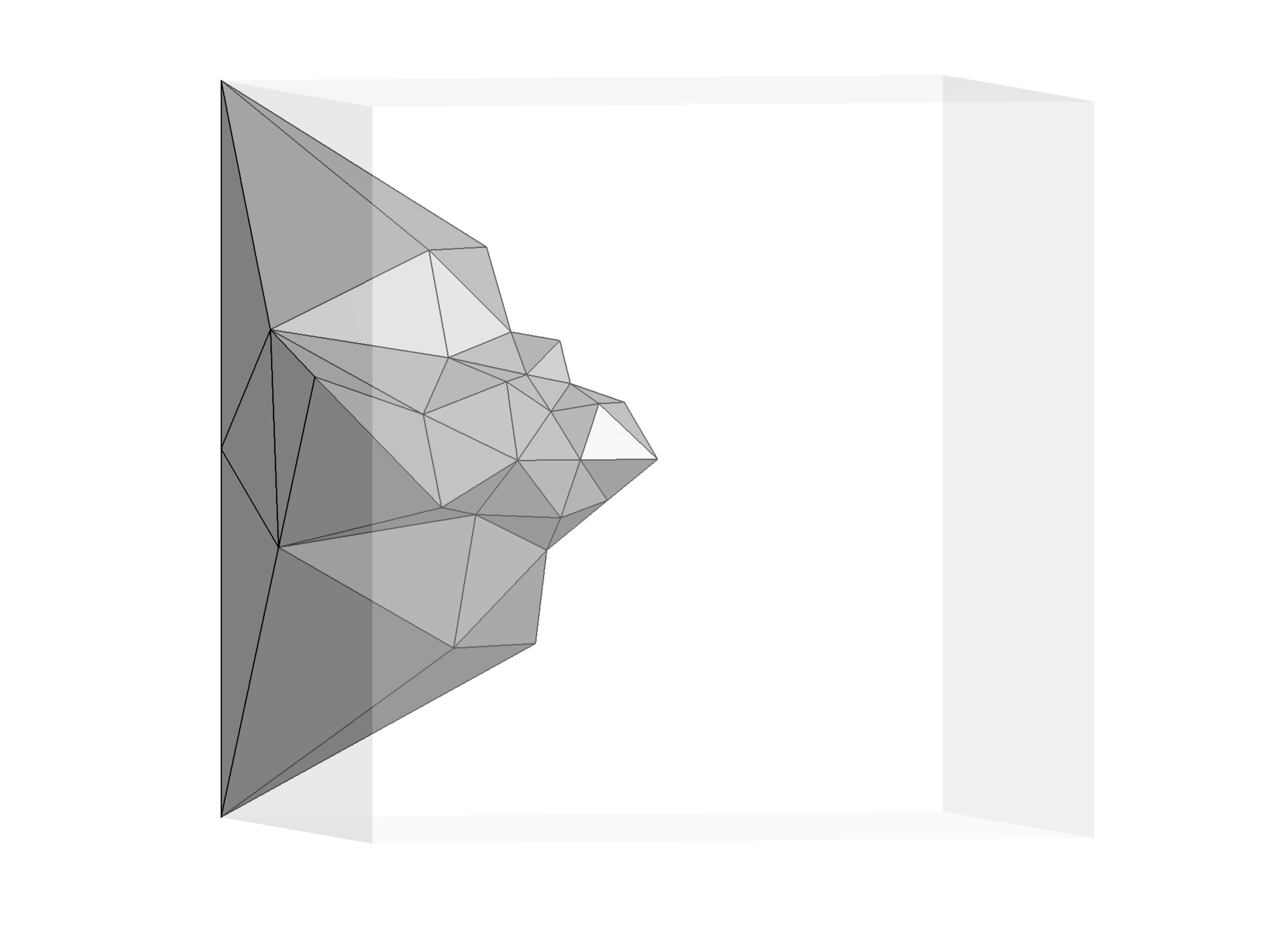}%
  \hspace{30pt}%
  \includegraphics[width=180pt, trim=398pt 217pt 333pt 163pt, clip]%
  {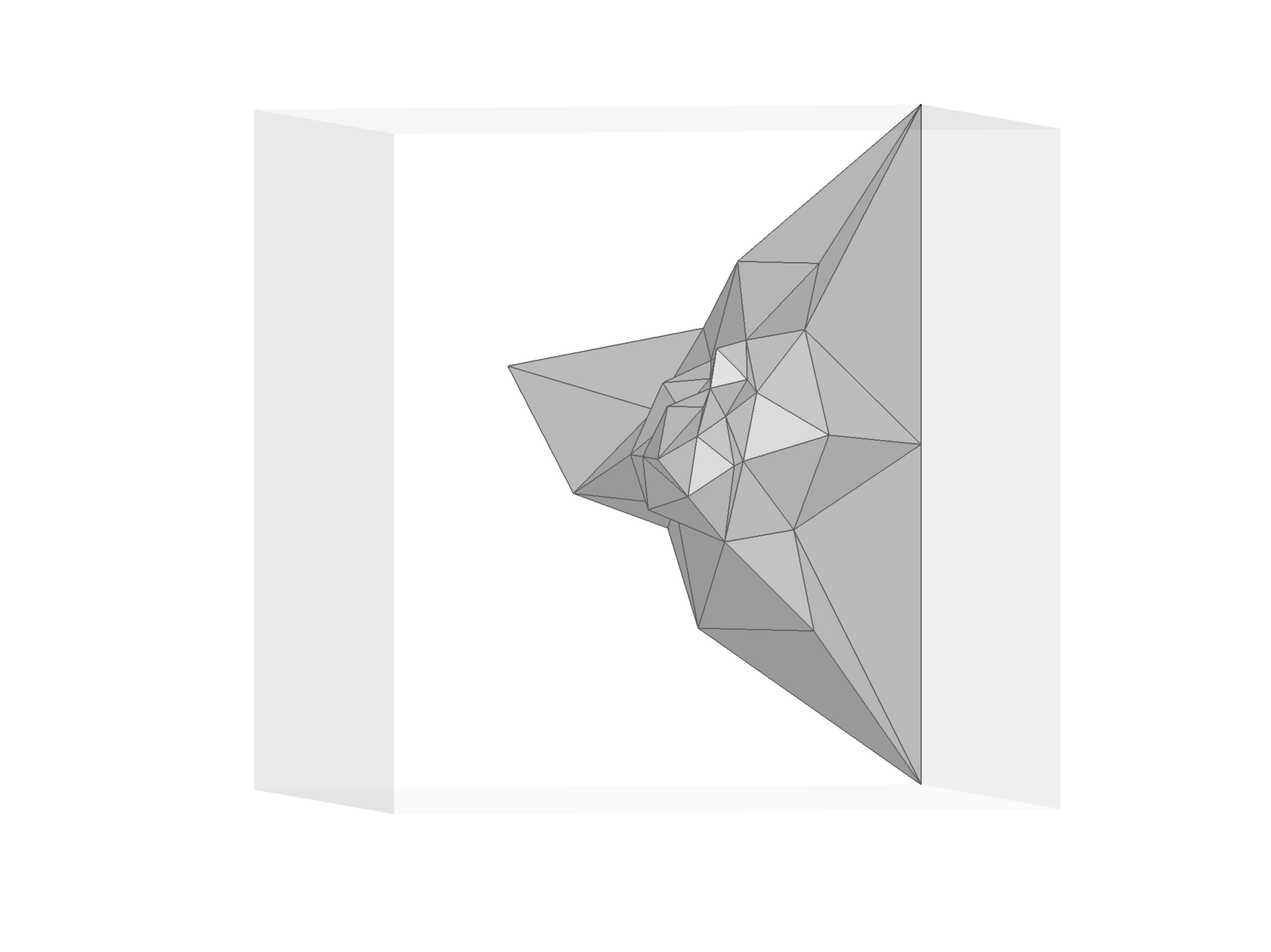}%
  \caption{Two views of a cutaway section of the acute triangulation
    of the cube.  The view at right is a $90$\textdegree\ rotation
    about the $z$-axis from the view at left.  On the left four of the
    triangles on the surface of the cube are visible.  These triangles
    appear on the back face of the cube in the view on the right.
    This cutaway is a collection of one of each of the $82$ distinct
    tetrahedra that are used in the acute triangulation.  The
    tetrahedra fit together to cover a symmetry region, and through
    rotations and reflections they can generate the full acute
    triangulation of the cube.}
  \label{fig:gentets}
\end{figure}

Because of the symmetries of this acute triangulation of the cube,
there are only $82$ distinct tetrahedra used in the $1370$-tetrahedron
acute triangulation of the cube. The cutaway views of the acute
triangulation of the cube in Fig.~\ref{fig:gentets} show just one of
each of these $82$ tetrahedra as they fit together to cover the
generating $1/24$th section of the cube. It is clear from
Fig.~\ref{fig:gentets} that many of the tetrahedra do not align with
the boundaries of the generating $1/24$th section of the cube.
Tetrahedra that intersect the boundaries of the generating region are
mapped onto themselves by one or more of the symmetries. In fact,
there are only $38$ tetrahedra that are interior to the generating
section of the cube. There are, of course, $24$ copies of each of
these tetrahedra in the final result. As far as the other tetrahedra
in the generating set are concerned, $35$ of them intersect one of the
planes bounding the region, $8$ of them intersect one of the main
diagonals of the cube, and $1$ intersects the $y$-axis. With some
thought about the symmetries involved, one can see that in the full
acute triangulation of the cube there are $12$ copies of each
tetrahedron that intersect a plane, $4$ copies of each tetrahedron
that intersect a main diagonal, and $6$ copies of the tetrahedron that
intersects the $y$-axis.

\section{Conclusions}
\label{sec:cnclsn}
We have demonstrated that there exists an acute triangulation of the
cube. The fact that the cube has an acute triangulation directly
implies that many other regions of $\RR^{3}$ also have acute
triangulations. In particular, one can reflect the acute triangulation
of the cube through one of its faces to get an acute triangulation of
a square prism twice as long as a cube with vertices that match on the
two square faces. By identifying the matching vertices with each other
one can get an acute triangulation of a periodic domain.
Alternatively, one can stack infinitely many of these objects together
to obtain an acute triangulation of an infinitely long square prism.
Using reflections and translations of this acute triangulation, one
can easily obtain an acute triangulation of an infinite slab in
$\RR^{3}$, or of all of $\RR^{3}$, as alternatives to the
constructions in~\cite{EpSuUn2004}. In fact, one can use translations
and reflections of an initial acute triangulation of the cube to
acutely triangulate any object in $\RR^{3}$ that can be tiled with
cubes. This can be used to create an infinite variety of acute
triangulations of the cube itself.

But does every polyhedron have a dihedral acute triangulation? This
remains an open question. Does every tetrahedron have a dihedral acute
triangulation? This question, too, remains open, and so far there is
still no nontrivial acute triangulation of the regular tetrahedron
that is known to the authors. It is likely that a computer-assisted
construction like the one discussed in this paper could be used to
obtain such an acute triangulation, but there may be more direct
methods. A directly constructive, perhaps simpler, acute triangulation
of the cube would also be of interest.

Another open problem is that of finding the smallest possible acute
triangulation of the cube, where size is measured in terms of the
number of tetrahedra. It may be that the 1370~tetrahedra acute
triangulation presented here is the smallest acute triangulation of
the cube possible, but the authors suspect this is not the case. The
analogous question in two dimensions---the smallest acute
triangulation of the square---has been answered; an acute
triangulation of the square requires at least eight
triangles~\cite{CaLo1980}.

\section*{Acknowledgement}
  The authors thank Edgar Ramos for helpful discussions.  The work of
  Anil N. Hirani and Evan VanderZee was supported by an NSF CAREER
  Award (Grant No. DMS-0645604). Vadim Zharnitsky was partially
  supported by NSF grant DMS 08-07897.

\bibliographystyle{acmurldoi}
\bibliography{wct}

\appendix

\section{Vertex Coordinates}
\label{sec:vtxcoord}

The coordinates of the vertices of an acute triangulation of the
cube are listed in Tables~\ref{table:vtxcoordsym1}
through~\ref{table:vtxcoordsym6}.  The full set of vertices is
the union of the vertices listed in the individual tables; the
tables separate the vertices according to the way they are
generated from the symmetries described in Sec.~\ref{sec:symmetry}.

% With article documentclass, separation of 15pt is needed
\settowidth{\dotlen}{.}
\begin{longtable}{| r @{.} l | r @{.} l | r @{.} l | p{15pt} | r @{.} l | r @{.} l | r @{.} l | p{15pt} | r @{.} l | r @{.} l | r @{.} l |}
$-0$ & $555$   &   $-0$ & $555$   &   $-0$ & $555$   & &   $-0$ & $555$   &   $-0$ & $555$   &   $-0$ & $555$   & &   $-0$ & $555$   &   $-0$ & $555$   &   $-0$ & $555$\kill
\caption{Coordinates of the $26$ vertices in the generating
  section of the cube.}
\label{table:vtxcoordsym1}\\
\cline{1-6} \cline{8-13} \cline{15-20}
\multicolumn{2}{|c|}{$x$}
    & \multicolumn{2}{c|}{$y$} & \multicolumn{2}{c|}{$z$} & &
\multicolumn{2}{c|}{$x$}
    & \multicolumn{2}{c|}{$y$} & \multicolumn{2}{c|}{$z$} & &
\multicolumn{2}{c|}{$x$}
    & \multicolumn{2}{c|}{$y$} & \multicolumn{2}{c|}{$z$}\\
\cline{1-6} \cline{8-13} \cline{15-20}
\multicolumn{1}{|r @{\hspace{\dotlen}}}{$-1$} &    &   \multicolumn{1}{r @{\hspace{\dotlen}}}{$-1$} &    &   \multicolumn{1}{r @{\hspace{\dotlen}}}{$-1$} &    & &   $-0$ & $152$   &   $-0$ & $472$   &   $-0$ & $152$   & &   $-0$ & $127$   &   $-0$ & $269$   &   $0$ & $127$ \\
% NOTE: The following line and similar line at the end of the
% \head section could be removed if we could guarantee that the
% page break will never occur between the \cline commands in the
% table and the subsequent row in the table.  It seems like
% the longtable package ought to guarantee this behavior, but
% experimental evidence proves it does not.  In a final version,
% a \newpage command could be inserted as necessary to force
% the page break to occur in the proper place
\cline{1-6} \cline{8-13} \cline{15-20}
\endfirsthead
\caption[]{(continued)}\\
\cline{1-6} \cline{8-13} \cline{15-20}
\multicolumn{2}{|c|}{$x$}
    & \multicolumn{2}{c|}{$y$} & \multicolumn{2}{c|}{$z$} & &
\multicolumn{2}{c|}{$x$}
    & \multicolumn{2}{c|}{$y$} & \multicolumn{2}{c|}{$z$} & &
\multicolumn{2}{c|}{$x$}
    & \multicolumn{2}{c|}{$y$} & \multicolumn{2}{c|}{$z$}\\
% NOTE: The following line is similar to line at end of \firsthead
% section -- see note there
\cline{1-6} \cline{8-13} \cline{15-20}
\endhead
\cline{1-6} \cline{8-13} \cline{15-20}
\endfoot
\cline{1-6} \cline{8-13} \cline{15-20}
$-0$ & $122$   &   $-0$ & $624$   &   $0$ & $122$   & &   $-0$ & $21$   &   $-0$ & $398$   &   $0$ & $099$   & &   \multicolumn{1}{r @{\hspace{\dotlen}}}{$0$} &    &   \multicolumn{1}{r @{\hspace{\dotlen}}}{$0$} &    &   \multicolumn{1}{r @{\hspace{\dotlen}}}{$0$} &  \\
\cline{1-6} \cline{8-13} \cline{15-20}
$-0$ & $399$   &   $-0$ & $598$   &   $0$ & $052$   & &   $-0$ & $224$   &   $-0$ & $316$   &   $0$ & $224$ & \multicolumn{7}{l}{$\quad$}\\
\cline{1-6} \cline{8-13}
\endlastfoot
% This line would have to be added back if lines in firsthead 
% and head were removed
%\cline{1-6} \cline{8-13} \cline{15-20}
$-0$ & $24$   &   \multicolumn{1}{r @{\hspace{\dotlen}}}{$-1$} &    &   $-0$ & $24$   & &   $-0$ & $27$   &   $-0$ & $523$   &   $0$ & $27$   & &   $-0$ & $158$   &   $-0$ & $263$   &   $-0$ & $027$ \\
\cline{1-6} \cline{8-13} \cline{15-20}\multicolumn{1}{|r @{\hspace{\dotlen}}}{$-1$} &    &   \multicolumn{1}{r @{\hspace{\dotlen}}}{$-1$} &    &   \multicolumn{1}{r @{\hspace{\dotlen}}}{$0$} &    & &   $-0$ & $254$   &   $-0$ & $254$   &   $-0$ & $254$   & &   $-0$ & $2$   &   $-0$ & $2$   &   $0$ & $2$ \\
\cline{1-6} \cline{8-13} \cline{15-20}$-0$ & $347$   &   \multicolumn{1}{r @{\hspace{\dotlen}}}{$-1$} &    &   $0$ & $347$   & &   $-0$ & $336$   &   $-0$ & $336$   &   $0$ & $336$   & &   $-0$ & $258$   &   $-0$ & $258$   &   $0$ & $115$ \\
\cline{1-6} \cline{8-13} \cline{15-20}\multicolumn{1}{|r @{\hspace{\dotlen}}}{$-1$} &    &   \multicolumn{1}{r @{\hspace{\dotlen}}}{$-1$} &    &   \multicolumn{1}{r @{\hspace{\dotlen}}}{$1$} &    & &   $-0$ & $325$   &   $-0$ & $325$   &   $-0$ & $02$   & &   $-0$ & $115$   &   $-0$ & $115$   &   $-0$ & $115$ \\
\cline{1-6} \cline{8-13} \cline{15-20}$-0$ & $517$   &   $-0$ & $517$   &   $-0$ & $23$   & &   $-0$ & $376$   &   $-0$ & $376$   &   $0$ & $18$   & &   \multicolumn{1}{r @{\hspace{\dotlen}}}{$0$} &    &   $-0$ & $214$   &   \multicolumn{1}{r @{\hspace{\dotlen}}}{$0$} &  \\
\cline{1-6} \cline{8-13} \cline{15-20}$-0$ & $559$   &   $-0$ & $559$   &   $0$ & $357$   & &   \multicolumn{1}{r @{\hspace{\dotlen}}}{$0$} &    &   $-0$ & $388$   &   \multicolumn{1}{r @{\hspace{\dotlen}}}{$0$} &    & &   $-0$ & $15$   &   $-0$ & $15$   &   $0$ & $061$ \\
\end{longtable}

\begin{longtable}{| r @{.} l | r @{.} l | r @{.} l | p{15pt} | r @{.} l | r @{.} l | r @{.} l | p{15pt} | r @{.} l | r @{.} l | r @{.} l |}
$-0$ & $555$   &   $-0$ & $555$   &   $-0$ & $555$   & &   $-0$ & $555$   &   $-0$ & $555$   &   $-0$ & $555$   & &   $-0$ & $555$   &   $-0$ & $555$   &   $-0$ & $555$\kill
\caption{Coordinates of the $15$ new vertices obtained by applying
  matrix $A_{1}$ to the vertices in Table~\ref{table:vtxcoordsym1}.}
\label{table:vtxcoordsym2}\\
\cline{1-6} \cline{8-13} \cline{15-20}
\multicolumn{2}{|c|}{$x$}
    & \multicolumn{2}{c|}{$y$} & \multicolumn{2}{c|}{$z$} & &
\multicolumn{2}{c|}{$x$}
    & \multicolumn{2}{c|}{$y$} & \multicolumn{2}{c|}{$z$} & &
\multicolumn{2}{c|}{$x$}
    & \multicolumn{2}{c|}{$y$} & \multicolumn{2}{c|}{$z$}\\
\cline{1-6} \cline{8-13} \cline{15-20}
\multicolumn{1}{|r @{\hspace{\dotlen}}}{$1$} &    &   \multicolumn{1}{r @{\hspace{\dotlen}}}{$-1$} &    &   \multicolumn{1}{r @{\hspace{\dotlen}}}{$1$} &    & &   $-0$ & $052$   &   $-0$ & $598$   &   $0$ & $399$   & &   $-0$ & $099$   &   $-0$ & $398$   &   $0$ & $21$ \\
% NOTE: The following line and similar line at the end of the
% \head section could be removed if we could guarantee that the
% page break will never occur between the \cline commands in the
% table and the subsequent row in the table.  It seems like
% the longtable package ought to guarantee this behavior, but
% experimental evidence proves it does not.  In a final version,
% a \newpage command could be inserted as necessary to force
% the page break to occur in the proper place
\cline{1-6} \cline{8-13} \cline{15-20}
\endfirsthead
\caption[]{(continued)}\\
\cline{1-6} \cline{8-13} \cline{15-20}
\multicolumn{2}{|c|}{$x$}
    & \multicolumn{2}{c|}{$y$} & \multicolumn{2}{c|}{$z$} & &
\multicolumn{2}{c|}{$x$}
    & \multicolumn{2}{c|}{$y$} & \multicolumn{2}{c|}{$z$} & &
\multicolumn{2}{c|}{$x$}
    & \multicolumn{2}{c|}{$y$} & \multicolumn{2}{c|}{$z$}\\
% NOTE: The following line is similar to line at end of \firsthead
% section -- see note there
\cline{1-6} \cline{8-13} \cline{15-20}
\endhead
\cline{1-6} \cline{8-13} \cline{15-20}
\endfoot
\cline{1-6} \cline{8-13} \cline{15-20}
$-0$ & $357$   &   $-0$ & $559$   &   $0$ & $559$   & &   $-0$ & $18$   &   $-0$ & $376$   &   $0$ & $376$   & &   $-0$ & $061$   &   $-0$ & $15$   &   $0$ & $15$ \\
\cline{1-6} \cline{8-13} \cline{15-20}
\endlastfoot
% This line would have to be added back if lines in firsthead 
% and head were removed
%\cline{1-6} \cline{8-13} \cline{15-20}
$0$ & $24$   &   \multicolumn{1}{r @{\hspace{\dotlen}}}{$-1$} &    &   $0$ & $24$   & &   $0$ & $152$   &   $-0$ & $472$   &   $0$ & $152$   & &   $0$ & $027$   &   $-0$ & $263$   &   $0$ & $158$ \\
\cline{1-6} \cline{8-13} \cline{15-20}\multicolumn{1}{|r @{\hspace{\dotlen}}}{$0$} &    &   \multicolumn{1}{r @{\hspace{\dotlen}}}{$-1$} &    &   \multicolumn{1}{r @{\hspace{\dotlen}}}{$1$} &    & &   $0$ & $254$   &   $-0$ & $254$   &   $0$ & $254$   & &   $-0$ & $115$   &   $-0$ & $258$   &   $0$ & $258$ \\
\cline{1-6} \cline{8-13} \cline{15-20}$0$ & $23$   &   $-0$ & $517$   &   $0$ & $517$   & &   $0$ & $02$   &   $-0$ & $325$   &   $0$ & $325$   & &   $0$ & $115$   &   $-0$ & $115$   &   $0$ & $115$ \\
\end{longtable}

\begin{longtable}{| r @{.} l | r @{.} l | r @{.} l | p{15pt} | r @{.} l | r @{.} l | r @{.} l | p{15pt} | r @{.} l | r @{.} l | r @{.} l |}
$-0$ & $555$   &   $-0$ & $555$   &   $-0$ & $555$   & &   $-0$ & $555$   &   $-0$ & $555$   &   $-0$ & $555$   & &   $-0$ & $555$   &   $-0$ & $555$   &   $-0$ & $555$\kill
\caption{Coordinates of the $28$ new vertices obtained by applying
  matrix $A_{2}$ to the vertices in Tables~\ref{table:vtxcoordsym1}
  and~\ref{table:vtxcoordsym2}.}
\label{table:vtxcoordsym3}\\
\cline{1-6} \cline{8-13} \cline{15-20}
\multicolumn{2}{|c|}{$x$}
    & \multicolumn{2}{c|}{$y$} & \multicolumn{2}{c|}{$z$} & &
\multicolumn{2}{c|}{$x$}
    & \multicolumn{2}{c|}{$y$} & \multicolumn{2}{c|}{$z$} & &
\multicolumn{2}{c|}{$x$}
    & \multicolumn{2}{c|}{$y$} & \multicolumn{2}{c|}{$z$}\\
\cline{1-6} \cline{8-13} \cline{15-20}
\multicolumn{1}{|r @{\hspace{\dotlen}}}{$0$} &    &   \multicolumn{1}{r @{\hspace{\dotlen}}}{$-1$} &    &   \multicolumn{1}{r @{\hspace{\dotlen}}}{$-1$} &    & &   $0$ & $18$   &   $-0$ & $376$   &   $-0$ & $376$   & &   $0$ & $517$   &   $-0$ & $517$   &   $0$ & $23$ \\
% NOTE: The following line and similar line at the end of the
% \head section could be removed if we could guarantee that the
% page break will never occur between the \cline commands in the
% table and the subsequent row in the table.  It seems like
% the longtable package ought to guarantee this behavior, but
% experimental evidence proves it does not.  In a final version,
% a \newpage command could be inserted as necessary to force
% the page break to occur in the proper place
\cline{1-6} \cline{8-13} \cline{15-20}
\endfirsthead
\caption[]{(continued)}\\
\cline{1-6} \cline{8-13} \cline{15-20}
\multicolumn{2}{|c|}{$x$}
    & \multicolumn{2}{c|}{$y$} & \multicolumn{2}{c|}{$z$} & &
\multicolumn{2}{c|}{$x$}
    & \multicolumn{2}{c|}{$y$} & \multicolumn{2}{c|}{$z$} & &
\multicolumn{2}{c|}{$x$}
    & \multicolumn{2}{c|}{$y$} & \multicolumn{2}{c|}{$z$}\\
% NOTE: The following line is similar to line at end of \firsthead
% section -- see note there
\cline{1-6} \cline{8-13} \cline{15-20}
\endhead
\cline{1-6} \cline{8-13} \cline{15-20}
\endfoot
\cline{1-6} \cline{8-13} \cline{15-20}
$0$ & $336$   &   $-0$ & $336$   &   $-0$ & $336$   & &   \multicolumn{1}{r @{\hspace{\dotlen}}}{$1$} &    &   \multicolumn{1}{r @{\hspace{\dotlen}}}{$-1$} &    &   \multicolumn{1}{r @{\hspace{\dotlen}}}{$0$} &    & &   $0$ & $15$   &   $-0$ & $15$   &   $-0$ & $061$ \\
\cline{1-6} \cline{8-13} \cline{15-20}
$-0$ & $02$   &   $-0$ & $325$   &   $-0$ & $325$ & \multicolumn{14}{l}{$\quad$}\\
\cline{1-6}
\endlastfoot
% This line would have to be added back if lines in firsthead 
% and head were removed
%\cline{1-6} \cline{8-13} \cline{15-20}
$0$ & $347$   &   \multicolumn{1}{r @{\hspace{\dotlen}}}{$-1$} &    &   $-0$ & $347$   & &   $0$ & $099$   &   $-0$ & $398$   &   $-0$ & $21$   & &   $0$ & $559$   &   $-0$ & $559$   &   $-0$ & $357$ \\
\cline{1-6} \cline{8-13} \cline{15-20}\multicolumn{1}{|r @{\hspace{\dotlen}}}{$1$} &    &   \multicolumn{1}{r @{\hspace{\dotlen}}}{$-1$} &    &   \multicolumn{1}{r @{\hspace{\dotlen}}}{$-1$} &    & &   $0$ & $224$   &   $-0$ & $316$   &   $-0$ & $224$   & &   $0$ & $399$   &   $-0$ & $598$   &   $-0$ & $052$ \\
\cline{1-6} \cline{8-13} \cline{15-20}$-0$ & $23$   &   $-0$ & $517$   &   $-0$ & $517$   & &   $0$ & $127$   &   $-0$ & $269$   &   $-0$ & $127$   & &   $0$ & $325$   &   $-0$ & $325$   &   $0$ & $02$ \\
\cline{1-6} \cline{8-13} \cline{15-20}$0$ & $357$   &   $-0$ & $559$   &   $-0$ & $559$   & &   $-0$ & $027$   &   $-0$ & $263$   &   $-0$ & $158$   & &   $0$ & $376$   &   $-0$ & $376$   &   $-0$ & $18$ \\
\cline{1-6} \cline{8-13} \cline{15-20}$0$ & $122$   &   $-0$ & $624$   &   $-0$ & $122$   & &   $0$ & $2$   &   $-0$ & $2$   &   $-0$ & $2$   & &   $0$ & $21$   &   $-0$ & $398$   &   $-0$ & $099$ \\
\cline{1-6} \cline{8-13} \cline{15-20}$0$ & $052$   &   $-0$ & $598$   &   $-0$ & $399$   & &   $0$ & $115$   &   $-0$ & $258$   &   $-0$ & $258$   & &   $0$ & $158$   &   $-0$ & $263$   &   $0$ & $027$ \\
\cline{1-6} \cline{8-13} \cline{15-20}$0$ & $27$   &   $-0$ & $523$   &   $-0$ & $27$   & &   $0$ & $061$   &   $-0$ & $15$   &   $-0$ & $15$   & &   $0$ & $258$   &   $-0$ & $258$   &   $-0$ & $115$ \\
\end{longtable}

\begin{longtable}{| r @{.} l | r @{.} l | r @{.} l | p{15pt} | r @{.} l | r @{.} l | r @{.} l | p{15pt} | r @{.} l | r @{.} l | r @{.} l |}
$-0$ & $555$   &   $-0$ & $555$   &   $-0$ & $555$   & &   $-0$ & $555$   &   $-0$ & $555$   &   $-0$ & $555$   & &   $-0$ & $555$   &   $-0$ & $555$   &   $-0$ & $555$\kill
\caption{Coordinates of the $68$ new vertices obtained by applying
  matrix $A_{3}$ to the vertices in Tables~\ref{table:vtxcoordsym1}
  through~\ref{table:vtxcoordsym3}.}
\label{table:vtxcoordsym4}\\
\cline{1-6} \cline{8-13} \cline{15-20}
\multicolumn{2}{|c|}{$x$}
    & \multicolumn{2}{c|}{$y$} & \multicolumn{2}{c|}{$z$} & &
\multicolumn{2}{c|}{$x$}
    & \multicolumn{2}{c|}{$y$} & \multicolumn{2}{c|}{$z$} & &
\multicolumn{2}{c|}{$x$}
    & \multicolumn{2}{c|}{$y$} & \multicolumn{2}{c|}{$z$}\\
\cline{1-6} \cline{8-13} \cline{15-20}
\multicolumn{1}{|r @{\hspace{\dotlen}}}{$1$} &    &   \multicolumn{1}{r @{\hspace{\dotlen}}}{$1$} &    &   \multicolumn{1}{r @{\hspace{\dotlen}}}{$-1$} &    & &   \multicolumn{1}{r @{\hspace{\dotlen}}}{$0$} &    &   $0$ & $214$   &   \multicolumn{1}{r @{\hspace{\dotlen}}}{$0$} &    & &   $-0$ & $052$   &   $0$ & $598$   &   $-0$ & $399$ \\
% NOTE: The following line and similar line at the end of the
% \head section could be removed if we could guarantee that the
% page break will never occur between the \cline commands in the
% table and the subsequent row in the table.  It seems like
% the longtable package ought to guarantee this behavior, but
% experimental evidence proves it does not.  In a final version,
% a \newpage command could be inserted as necessary to force
% the page break to occur in the proper place
\cline{1-6} \cline{8-13} \cline{15-20}
\endfirsthead
\caption[]{(continued)}\\
\cline{1-6} \cline{8-13} \cline{15-20}
\multicolumn{2}{|c|}{$x$}
    & \multicolumn{2}{c|}{$y$} & \multicolumn{2}{c|}{$z$} & &
\multicolumn{2}{c|}{$x$}
    & \multicolumn{2}{c|}{$y$} & \multicolumn{2}{c|}{$z$} & &
\multicolumn{2}{c|}{$x$}
    & \multicolumn{2}{c|}{$y$} & \multicolumn{2}{c|}{$z$}\\
% NOTE: The following line is similar to line at end of \firsthead
% section -- see note there
\cline{1-6} \cline{8-13} \cline{15-20}
\endhead
\cline{1-6} \cline{8-13} \cline{15-20}
\endfoot
\cline{1-6} \cline{8-13} \cline{15-20}
$0$ & $258$   &   $0$ & $258$   &   $0$ & $115$   & &   $-0$ & $357$   &   $0$ & $559$   &   $-0$ & $559$   & &   $-0$ & $15$   &   $0$ & $15$   &   $-0$ & $061$ \\
\cline{1-6} \cline{8-13} \cline{15-20}
$0$ & $115$   &   $0$ & $115$   &   $-0$ & $115$   & &   $-0$ & $122$   &   $0$ & $624$   &   $-0$ & $122$ & \multicolumn{7}{l}{$\quad$}\\
\cline{1-6} \cline{8-13}
\endlastfoot
% This line would have to be added back if lines in firsthead 
% and head were removed
%\cline{1-6} \cline{8-13} \cline{15-20}
$0$ & $24$   &   \multicolumn{1}{r @{\hspace{\dotlen}}}{$1$} &    &   $-0$ & $24$   & &   $0$ & $15$   &   $0$ & $15$   &   $0$ & $061$   & &   $-0$ & $27$   &   $0$ & $523$   &   $-0$ & $27$ \\
\cline{1-6} \cline{8-13} \cline{15-20}\multicolumn{1}{|r @{\hspace{\dotlen}}}{$1$} &    &   \multicolumn{1}{r @{\hspace{\dotlen}}}{$1$} &    &   \multicolumn{1}{r @{\hspace{\dotlen}}}{$0$} &    & &   \multicolumn{1}{r @{\hspace{\dotlen}}}{$-1$} &    &   \multicolumn{1}{r @{\hspace{\dotlen}}}{$1$} &    &   \multicolumn{1}{r @{\hspace{\dotlen}}}{$1$} &    & &   $-0$ & $336$   &   $0$ & $336$   &   $-0$ & $336$ \\
\cline{1-6} \cline{8-13} \cline{15-20}$0$ & $347$   &   \multicolumn{1}{r @{\hspace{\dotlen}}}{$1$} &    &   $0$ & $347$   & &   $-0$ & $24$   &   \multicolumn{1}{r @{\hspace{\dotlen}}}{$1$} &    &   $0$ & $24$   & &   $0$ & $02$   &   $0$ & $325$   &   $-0$ & $325$ \\
\cline{1-6} \cline{8-13} \cline{15-20}\multicolumn{1}{|r @{\hspace{\dotlen}}}{$1$} &    &   \multicolumn{1}{r @{\hspace{\dotlen}}}{$1$} &    &   \multicolumn{1}{r @{\hspace{\dotlen}}}{$1$} &    & &   \multicolumn{1}{r @{\hspace{\dotlen}}}{$0$} &    &   \multicolumn{1}{r @{\hspace{\dotlen}}}{$1$} &    &   \multicolumn{1}{r @{\hspace{\dotlen}}}{$1$} &    & &   $-0$ & $18$   &   $0$ & $376$   &   $-0$ & $376$ \\
\cline{1-6} \cline{8-13} \cline{15-20}$0$ & $517$   &   $0$ & $517$   &   $-0$ & $23$   & &   $-0$ & $23$   &   $0$ & $517$   &   $0$ & $517$   & &   $-0$ & $099$   &   $0$ & $398$   &   $-0$ & $21$ \\
\cline{1-6} \cline{8-13} \cline{15-20}$0$ & $559$   &   $0$ & $559$   &   $0$ & $357$   & &   $0$ & $357$   &   $0$ & $559$   &   $0$ & $559$   & &   $-0$ & $224$   &   $0$ & $316$   &   $-0$ & $224$ \\
\cline{1-6} \cline{8-13} \cline{15-20}$0$ & $122$   &   $0$ & $624$   &   $0$ & $122$   & &   $0$ & $052$   &   $0$ & $598$   &   $0$ & $399$   & &   $-0$ & $127$   &   $0$ & $269$   &   $-0$ & $127$ \\
\cline{1-6} \cline{8-13} \cline{15-20}$0$ & $399$   &   $0$ & $598$   &   $0$ & $052$   & &   $-0$ & $152$   &   $0$ & $472$   &   $0$ & $152$   & &   $0$ & $027$   &   $0$ & $263$   &   $-0$ & $158$ \\
\cline{1-6} \cline{8-13} \cline{15-20}$0$ & $152$   &   $0$ & $472$   &   $-0$ & $152$   & &   $-0$ & $254$   &   $0$ & $254$   &   $0$ & $254$   & &   $-0$ & $2$   &   $0$ & $2$   &   $-0$ & $2$ \\
\cline{1-6} \cline{8-13} \cline{15-20}$0$ & $27$   &   $0$ & $523$   &   $0$ & $27$   & &   $-0$ & $02$   &   $0$ & $325$   &   $0$ & $325$   & &   $-0$ & $115$   &   $0$ & $258$   &   $-0$ & $258$ \\
\cline{1-6} \cline{8-13} \cline{15-20}$0$ & $254$   &   $0$ & $254$   &   $-0$ & $254$   & &   $0$ & $18$   &   $0$ & $376$   &   $0$ & $376$   & &   $-0$ & $061$   &   $0$ & $15$   &   $-0$ & $15$ \\
\cline{1-6} \cline{8-13} \cline{15-20}$0$ & $336$   &   $0$ & $336$   &   $0$ & $336$   & &   $0$ & $099$   &   $0$ & $398$   &   $0$ & $21$   & &   \multicolumn{1}{r @{\hspace{\dotlen}}}{$-1$} &    &   \multicolumn{1}{r @{\hspace{\dotlen}}}{$1$} &    &   \multicolumn{1}{r @{\hspace{\dotlen}}}{$0$} &  \\
\cline{1-6} \cline{8-13} \cline{15-20}$0$ & $325$   &   $0$ & $325$   &   $-0$ & $02$   & &   $-0$ & $027$   &   $0$ & $263$   &   $0$ & $158$   & &   $-0$ & $517$   &   $0$ & $517$   &   $0$ & $23$ \\
\cline{1-6} \cline{8-13} \cline{15-20}$0$ & $376$   &   $0$ & $376$   &   $0$ & $18$   & &   $0$ & $115$   &   $0$ & $258$   &   $0$ & $258$   & &   $-0$ & $559$   &   $0$ & $559$   &   $-0$ & $357$ \\
\cline{1-6} \cline{8-13} \cline{15-20}\multicolumn{1}{|r @{\hspace{\dotlen}}}{$0$} &    &   $0$ & $388$   &   \multicolumn{1}{r @{\hspace{\dotlen}}}{$0$} &    & &   $-0$ & $115$   &   $0$ & $115$   &   $0$ & $115$   & &   $-0$ & $399$   &   $0$ & $598$   &   $-0$ & $052$ \\
\cline{1-6} \cline{8-13} \cline{15-20}$0$ & $21$   &   $0$ & $398$   &   $0$ & $099$   & &   $0$ & $061$   &   $0$ & $15$   &   $0$ & $15$   & &   $-0$ & $325$   &   $0$ & $325$   &   $0$ & $02$ \\
\cline{1-6} \cline{8-13} \cline{15-20}$0$ & $224$   &   $0$ & $316$   &   $0$ & $224$   & &   \multicolumn{1}{r @{\hspace{\dotlen}}}{$0$} &    &   \multicolumn{1}{r @{\hspace{\dotlen}}}{$1$} &    &   \multicolumn{1}{r @{\hspace{\dotlen}}}{$-1$} &    & &   $-0$ & $376$   &   $0$ & $376$   &   $-0$ & $18$ \\
\cline{1-6} \cline{8-13} \cline{15-20}$0$ & $127$   &   $0$ & $269$   &   $0$ & $127$   & &   $-0$ & $347$   &   \multicolumn{1}{r @{\hspace{\dotlen}}}{$1$} &    &   $-0$ & $347$   & &   $-0$ & $21$   &   $0$ & $398$   &   $-0$ & $099$ \\
\cline{1-6} \cline{8-13} \cline{15-20}$0$ & $158$   &   $0$ & $263$   &   $-0$ & $027$   & &   \multicolumn{1}{r @{\hspace{\dotlen}}}{$-1$} &    &   \multicolumn{1}{r @{\hspace{\dotlen}}}{$1$} &    &   \multicolumn{1}{r @{\hspace{\dotlen}}}{$-1$} &    & &   $-0$ & $158$   &   $0$ & $263$   &   $0$ & $027$ \\
\cline{1-6} \cline{8-13} \cline{15-20}$0$ & $2$   &   $0$ & $2$   &   $0$ & $2$   & &   $0$ & $23$   &   $0$ & $517$   &   $-0$ & $517$   & &   $-0$ & $258$   &   $0$ & $258$   &   $-0$ & $115$ \\
\end{longtable}

\begin{longtable}{| r @{.} l | r @{.} l | r @{.} l | p{15pt} | r @{.} l | r @{.} l | r @{.} l | p{15pt} | r @{.} l | r @{.} l | r @{.} l |}
$-0$ & $555$   &   $-0$ & $555$   &   $-0$ & $555$   & &   $-0$ & $555$   &   $-0$ & $555$   &   $-0$ & $555$   & &   $-0$ & $555$   &   $-0$ & $555$   &   $-0$ & $555$\kill
\caption{Coordinates of the $84$ new vertices obtained by applying
  matrix $A_{4}$ to the vertices in Tables~\ref{table:vtxcoordsym1}
  through~\ref{table:vtxcoordsym4}.}
\label{table:vtxcoordsym5}\\
\cline{1-6} \cline{8-13} \cline{15-20}
\multicolumn{2}{|c|}{$x$}
    & \multicolumn{2}{c|}{$y$} & \multicolumn{2}{c|}{$z$} & &
\multicolumn{2}{c|}{$x$}
    & \multicolumn{2}{c|}{$y$} & \multicolumn{2}{c|}{$z$} & &
\multicolumn{2}{c|}{$x$}
    & \multicolumn{2}{c|}{$y$} & \multicolumn{2}{c|}{$z$}\\
\cline{1-6} \cline{8-13} \cline{15-20}
\multicolumn{1}{|r @{\hspace{\dotlen}}}{$-1$} &    &   $-0$ & $24$   &   $-0$ & $24$   & &   $-0$ & $398$   &   $-0$ & $21$   &   $0$ & $099$   & &   $0$ & $269$   &   $0$ & $127$   &   $0$ & $127$ \\
% NOTE: The following line and similar line at the end of the
% \head section could be removed if we could guarantee that the
% page break will never occur between the \cline commands in the
% table and the subsequent row in the table.  It seems like
% the longtable package ought to guarantee this behavior, but
% experimental evidence proves it does not.  In a final version,
% a \newpage command could be inserted as necessary to force
% the page break to occur in the proper place
\cline{1-6} \cline{8-13} \cline{15-20}
\endfirsthead
\caption[]{(continued)}\\
\cline{1-6} \cline{8-13} \cline{15-20}
\multicolumn{2}{|c|}{$x$}
    & \multicolumn{2}{c|}{$y$} & \multicolumn{2}{c|}{$z$} & &
\multicolumn{2}{c|}{$x$}
    & \multicolumn{2}{c|}{$y$} & \multicolumn{2}{c|}{$z$} & &
\multicolumn{2}{c|}{$x$}
    & \multicolumn{2}{c|}{$y$} & \multicolumn{2}{c|}{$z$}\\
% NOTE: The following line is similar to line at end of \firsthead
% section -- see note there
\cline{1-6} \cline{8-13} \cline{15-20}
\endhead
\cline{1-6} \cline{8-13} \cline{15-20}
\endfoot
\cline{1-6} \cline{8-13} \cline{15-20}
$-0$ & $523$   &   $-0$ & $27$   &   $0$ & $27$   & &   $0$ & $316$   &   $0$ & $224$   &   $0$ & $224$   & &   $0$ & $15$   &   $-0$ & $061$   &   $-0$ & $15$ \\
\cline{1-6} \cline{8-13} \cline{15-20}
\endlastfoot
% This line would have to be added back if lines in firsthead 
% and head were removed
%\cline{1-6} \cline{8-13} \cline{15-20}
\multicolumn{1}{|r @{\hspace{\dotlen}}}{$-1$} &    &   \multicolumn{1}{r @{\hspace{\dotlen}}}{$0$} &    &   \multicolumn{1}{r @{\hspace{\dotlen}}}{$-1$} &    & &   $-0$ & $316$   &   $-0$ & $224$   &   $0$ & $224$   & &   $0$ & $263$   &   $-0$ & $027$   &   $0$ & $158$ \\
\cline{1-6} \cline{8-13} \cline{15-20}\multicolumn{1}{|r @{\hspace{\dotlen}}}{$-1$} &    &   $0$ & $347$   &   $-0$ & $347$   & &   $-0$ & $269$   &   $-0$ & $127$   &   $0$ & $127$   & &   $0$ & $258$   &   $0$ & $115$   &   $0$ & $258$ \\
\cline{1-6} \cline{8-13} \cline{15-20}$-0$ & $517$   &   $-0$ & $23$   &   $-0$ & $517$   & &   $-0$ & $263$   &   $-0$ & $158$   &   $-0$ & $027$   & &   $0$ & $214$   &   \multicolumn{1}{r @{\hspace{\dotlen}}}{$0$} &    &   \multicolumn{1}{r @{\hspace{\dotlen}}}{$0$} &  \\
\cline{1-6} \cline{8-13} \cline{15-20}$-0$ & $559$   &   $0$ & $357$   &   $-0$ & $559$   & &   \multicolumn{1}{r @{\hspace{\dotlen}}}{$-1$} &    &   \multicolumn{1}{r @{\hspace{\dotlen}}}{$0$} &    &   \multicolumn{1}{r @{\hspace{\dotlen}}}{$1$} &    & &   $0$ & $15$   &   $0$ & $061$   &   $0$ & $15$ \\
\cline{1-6} \cline{8-13} \cline{15-20}$-0$ & $624$   &   $0$ & $122$   &   $-0$ & $122$   & &   $-0$ & $517$   &   $0$ & $23$   &   $0$ & $517$   & &   \multicolumn{1}{r @{\hspace{\dotlen}}}{$1$} &    &   $0$ & $24$   &   $-0$ & $24$ \\
\cline{1-6} \cline{8-13} \cline{15-20}$-0$ & $598$   &   $0$ & $052$   &   $-0$ & $399$   & &   $-0$ & $559$   &   $-0$ & $357$   &   $0$ & $559$   & &   $0$ & $598$   &   $0$ & $399$   &   $0$ & $052$ \\
\cline{1-6} \cline{8-13} \cline{15-20}$-0$ & $472$   &   $-0$ & $152$   &   $-0$ & $152$   & &   $-0$ & $598$   &   $-0$ & $052$   &   $0$ & $399$   & &   $0$ & $472$   &   $0$ & $152$   &   $-0$ & $152$ \\
\cline{1-6} \cline{8-13} \cline{15-20}$-0$ & $523$   &   $0$ & $27$   &   $-0$ & $27$   & &   $-0$ & $325$   &   $0$ & $02$   &   $0$ & $325$   & &   $0$ & $398$   &   $0$ & $21$   &   $0$ & $099$ \\
\cline{1-6} \cline{8-13} \cline{15-20}$-0$ & $325$   &   $-0$ & $02$   &   $-0$ & $325$   & &   $-0$ & $376$   &   $-0$ & $18$   &   $0$ & $376$   & &   $0$ & $263$   &   $0$ & $158$   &   $-0$ & $027$ \\
\cline{1-6} \cline{8-13} \cline{15-20}$-0$ & $376$   &   $0$ & $18$   &   $-0$ & $376$   & &   $-0$ & $398$   &   $-0$ & $099$   &   $0$ & $21$   & &   \multicolumn{1}{r @{\hspace{\dotlen}}}{$1$} &    &   $-0$ & $347$   &   $-0$ & $347$ \\
\cline{1-6} \cline{8-13} \cline{15-20}$-0$ & $388$   &   \multicolumn{1}{r @{\hspace{\dotlen}}}{$0$} &    &   \multicolumn{1}{r @{\hspace{\dotlen}}}{$0$} &    & &   $-0$ & $263$   &   $0$ & $027$   &   $0$ & $158$   & &   $0$ & $624$   &   $-0$ & $122$   &   $-0$ & $122$ \\
\cline{1-6} \cline{8-13} \cline{15-20}$-0$ & $398$   &   $0$ & $099$   &   $-0$ & $21$   & &   $-0$ & $258$   &   $-0$ & $115$   &   $0$ & $258$   & &   $0$ & $598$   &   $-0$ & $399$   &   $-0$ & $052$ \\
\cline{1-6} \cline{8-13} \cline{15-20}$-0$ & $316$   &   $0$ & $224$   &   $-0$ & $224$   & &   $-0$ & $15$   &   $-0$ & $061$   &   $0$ & $15$   & &   $0$ & $523$   &   $-0$ & $27$   &   $-0$ & $27$ \\
\cline{1-6} \cline{8-13} \cline{15-20}$-0$ & $269$   &   $0$ & $127$   &   $-0$ & $127$   & &   \multicolumn{1}{r @{\hspace{\dotlen}}}{$1$} &    &   $-0$ & $24$   &   $0$ & $24$   & &   $0$ & $398$   &   $-0$ & $21$   &   $-0$ & $099$ \\
\cline{1-6} \cline{8-13} \cline{15-20}$-0$ & $263$   &   $-0$ & $027$   &   $-0$ & $158$   & &   \multicolumn{1}{r @{\hspace{\dotlen}}}{$1$} &    &   \multicolumn{1}{r @{\hspace{\dotlen}}}{$0$} &    &   \multicolumn{1}{r @{\hspace{\dotlen}}}{$1$} &    & &   $0$ & $316$   &   $-0$ & $224$   &   $-0$ & $224$ \\
\cline{1-6} \cline{8-13} \cline{15-20}$-0$ & $258$   &   $0$ & $115$   &   $-0$ & $258$   & &   \multicolumn{1}{r @{\hspace{\dotlen}}}{$1$} &    &   $0$ & $347$   &   $0$ & $347$   & &   $0$ & $269$   &   $-0$ & $127$   &   $-0$ & $127$ \\
\cline{1-6} \cline{8-13} \cline{15-20}$-0$ & $214$   &   \multicolumn{1}{r @{\hspace{\dotlen}}}{$0$} &    &   \multicolumn{1}{r @{\hspace{\dotlen}}}{$0$} &    & &   $0$ & $517$   &   $-0$ & $23$   &   $0$ & $517$   & &   $0$ & $263$   &   $-0$ & $158$   &   $0$ & $027$ \\
\cline{1-6} \cline{8-13} \cline{15-20}$-0$ & $15$   &   $0$ & $061$   &   $-0$ & $15$   & &   $0$ & $559$   &   $0$ & $357$   &   $0$ & $559$   & &   \multicolumn{1}{r @{\hspace{\dotlen}}}{$1$} &    &   \multicolumn{1}{r @{\hspace{\dotlen}}}{$0$} &    &   \multicolumn{1}{r @{\hspace{\dotlen}}}{$-1$} &  \\
\cline{1-6} \cline{8-13} \cline{15-20}\multicolumn{1}{|r @{\hspace{\dotlen}}}{$-1$} &    &   $0$ & $24$   &   $0$ & $24$   & &   $0$ & $624$   &   $0$ & $122$   &   $0$ & $122$   & &   $0$ & $517$   &   $0$ & $23$   &   $-0$ & $517$ \\
\cline{1-6} \cline{8-13} \cline{15-20}$-0$ & $598$   &   $0$ & $399$   &   $-0$ & $052$   & &   $0$ & $598$   &   $0$ & $052$   &   $0$ & $399$   & &   $0$ & $559$   &   $-0$ & $357$   &   $-0$ & $559$ \\
\cline{1-6} \cline{8-13} \cline{15-20}$-0$ & $472$   &   $0$ & $152$   &   $0$ & $152$   & &   $0$ & $472$   &   $-0$ & $152$   &   $0$ & $152$   & &   $0$ & $598$   &   $-0$ & $052$   &   $-0$ & $399$ \\
\cline{1-6} \cline{8-13} \cline{15-20}$-0$ & $398$   &   $0$ & $21$   &   $-0$ & $099$   & &   $0$ & $523$   &   $0$ & $27$   &   $0$ & $27$   & &   $0$ & $325$   &   $0$ & $02$   &   $-0$ & $325$ \\
\cline{1-6} \cline{8-13} \cline{15-20}$-0$ & $263$   &   $0$ & $158$   &   $0$ & $027$   & &   $0$ & $325$   &   $-0$ & $02$   &   $0$ & $325$   & &   $0$ & $376$   &   $-0$ & $18$   &   $-0$ & $376$ \\
\cline{1-6} \cline{8-13} \cline{15-20}\multicolumn{1}{|r @{\hspace{\dotlen}}}{$-1$} &    &   $-0$ & $347$   &   $0$ & $347$   & &   $0$ & $376$   &   $0$ & $18$   &   $0$ & $376$   & &   $0$ & $398$   &   $-0$ & $099$   &   $-0$ & $21$ \\
\cline{1-6} \cline{8-13} \cline{15-20}$-0$ & $624$   &   $-0$ & $122$   &   $0$ & $122$   & &   $0$ & $388$   &   \multicolumn{1}{r @{\hspace{\dotlen}}}{$0$} &    &   \multicolumn{1}{r @{\hspace{\dotlen}}}{$0$} &    & &   $0$ & $263$   &   $0$ & $027$   &   $-0$ & $158$ \\
\cline{1-6} \cline{8-13} \cline{15-20}$-0$ & $598$   &   $-0$ & $399$   &   $0$ & $052$   & &   $0$ & $398$   &   $0$ & $099$   &   $0$ & $21$   & &   $0$ & $258$   &   $-0$ & $115$   &   $-0$ & $258$ \\
\end{longtable}

\newpage

\begin{longtable}{| r @{.} l | r @{.} l | r @{.} l | p{15pt} | r @{.} l | r @{.} l | r @{.} l | p{15pt} | r @{.} l | r @{.} l | r @{.} l |}
$-0$ & $555$   &   $-0$ & $555$   &   $-0$ & $555$   & &   $-0$ & $555$   &   $-0$ & $555$   &   $-0$ & $555$   & &   $-0$ & $555$   &   $-0$ & $555$   &   $-0$ & $555$\kill
\caption{Coordinates of the $56$ new vertices obtained by applying
  matrix $A_{4}^{2}$ to the vertices in Tables~\ref{table:vtxcoordsym1}
  through~\ref{table:vtxcoordsym4}.  Vertices that would be duplicated
  in Table~\ref{table:vtxcoordsym5} have also been removed.}
\label{table:vtxcoordsym6}\\
\cline{1-6} \cline{8-13} \cline{15-20}
\multicolumn{2}{|c|}{$x$}
    & \multicolumn{2}{c|}{$y$} & \multicolumn{2}{c|}{$z$} & &
\multicolumn{2}{c|}{$x$}
    & \multicolumn{2}{c|}{$y$} & \multicolumn{2}{c|}{$z$} & &
\multicolumn{2}{c|}{$x$}
    & \multicolumn{2}{c|}{$y$} & \multicolumn{2}{c|}{$z$}\\
\cline{1-6} \cline{8-13} \cline{15-20}
$-0$ & $24$   &   $-0$ & $24$   &   \multicolumn{1}{r @{\hspace{\dotlen}}}{$-1$} &    & &   $-0$ & $399$   &   $0$ & $052$   &   $-0$ & $598$   & &   $-0$ & $027$   &   $0$ & $158$   &   $0$ & $263$ \\
% NOTE: The following line and similar line at the end of the
% \head section could be removed if we could guarantee that the
% page break will never occur between the \cline commands in the
% table and the subsequent row in the table.  It seems like
% the longtable package ought to guarantee this behavior, but
% experimental evidence proves it does not.  In a final version,
% a \newpage command could be inserted as necessary to force
% the page break to occur in the proper place
\cline{1-6} \cline{8-13} \cline{15-20}
\endfirsthead
\caption[]{(continued)}\\
\cline{1-6} \cline{8-13} \cline{15-20}
\multicolumn{2}{|c|}{$x$}
    & \multicolumn{2}{c|}{$y$} & \multicolumn{2}{c|}{$z$} & &
\multicolumn{2}{c|}{$x$}
    & \multicolumn{2}{c|}{$y$} & \multicolumn{2}{c|}{$z$} & &
\multicolumn{2}{c|}{$x$}
    & \multicolumn{2}{c|}{$y$} & \multicolumn{2}{c|}{$z$}\\
% NOTE: The following line is similar to line at end of \firsthead
% section -- see note there
\cline{1-6} \cline{8-13} \cline{15-20}
\endhead
\cline{1-6} \cline{8-13} \cline{15-20}
\endfoot
\cline{1-6} \cline{8-13} \cline{15-20}
$-0$ & $347$   &   $0$ & $347$   &   \multicolumn{1}{r @{\hspace{\dotlen}}}{$-1$} &    & &   $0$ & $224$   &   $0$ & $224$   &   $0$ & $316$   & &   $0$ & $027$   &   $-0$ & $158$   &   $0$ & $263$ \\
\cline{1-6} \cline{8-13} \cline{15-20}
$-0$ & $122$   &   $0$ & $122$   &   $-0$ & $624$   & &   $0$ & $127$   &   $0$ & $127$   &   $0$ & $269$ & \multicolumn{7}{l}{$\quad$}\\
\cline{1-6} \cline{8-13}
\endlastfoot
% This line would have to be added back if lines in firsthead 
% and head were removed
%\cline{1-6} \cline{8-13} \cline{15-20}
$0$ & $347$   &   $-0$ & $347$   &   \multicolumn{1}{r @{\hspace{\dotlen}}}{$-1$} &    & &   $-0$ & $27$   &   $0$ & $27$   &   $-0$ & $523$   & &   \multicolumn{1}{r @{\hspace{\dotlen}}}{$0$} &    &   \multicolumn{1}{r @{\hspace{\dotlen}}}{$0$} &    &   $0$ & $214$ \\
\cline{1-6} \cline{8-13} \cline{15-20}$0$ & $122$   &   $-0$ & $122$   &   $-0$ & $624$   & &   $-0$ & $21$   &   $0$ & $099$   &   $-0$ & $398$   & &   $0$ & $24$   &   $-0$ & $24$   &   \multicolumn{1}{r @{\hspace{\dotlen}}}{$1$} &  \\
\cline{1-6} \cline{8-13} \cline{15-20}$0$ & $052$   &   $-0$ & $399$   &   $-0$ & $598$   & &   $-0$ & $224$   &   $0$ & $224$   &   $-0$ & $316$   & &   $0$ & $399$   &   $0$ & $052$   &   $0$ & $598$ \\
\cline{1-6} \cline{8-13} \cline{15-20}$-0$ & $152$   &   $-0$ & $152$   &   $-0$ & $472$   & &   $-0$ & $127$   &   $0$ & $127$   &   $-0$ & $269$   & &   $0$ & $152$   &   $-0$ & $152$   &   $0$ & $472$ \\
\cline{1-6} \cline{8-13} \cline{15-20}$0$ & $27$   &   $-0$ & $27$   &   $-0$ & $523$   & &   $-0$ & $158$   &   $-0$ & $027$   &   $-0$ & $263$   & &   $0$ & $21$   &   $0$ & $099$   &   $0$ & $398$ \\
\cline{1-6} \cline{8-13} \cline{15-20}\multicolumn{1}{|r @{\hspace{\dotlen}}}{$0$} &    &   \multicolumn{1}{r @{\hspace{\dotlen}}}{$0$} &    &   $-0$ & $388$   & &   $-0$ & $052$   &   $0$ & $399$   &   $-0$ & $598$   & &   $0$ & $158$   &   $-0$ & $027$   &   $0$ & $263$ \\
\cline{1-6} \cline{8-13} \cline{15-20}$0$ & $099$   &   $-0$ & $21$   &   $-0$ & $398$   & &   $-0$ & $099$   &   $0$ & $21$   &   $-0$ & $398$   & &   $-0$ & $347$   &   $-0$ & $347$   &   \multicolumn{1}{r @{\hspace{\dotlen}}}{$1$} &  \\
\cline{1-6} \cline{8-13} \cline{15-20}$0$ & $224$   &   $-0$ & $224$   &   $-0$ & $316$   & &   $0$ & $027$   &   $0$ & $158$   &   $-0$ & $263$   & &   $-0$ & $122$   &   $-0$ & $122$   &   $0$ & $624$ \\
\cline{1-6} \cline{8-13} \cline{15-20}$0$ & $127$   &   $-0$ & $127$   &   $-0$ & $269$   & &   $-0$ & $24$   &   $0$ & $24$   &   \multicolumn{1}{r @{\hspace{\dotlen}}}{$1$} &    & &   $-0$ & $399$   &   $-0$ & $052$   &   $0$ & $598$ \\
\cline{1-6} \cline{8-13} \cline{15-20}$-0$ & $027$   &   $-0$ & $158$   &   $-0$ & $263$   & &   $0$ & $347$   &   $0$ & $347$   &   \multicolumn{1}{r @{\hspace{\dotlen}}}{$1$} &    & &   $-0$ & $27$   &   $-0$ & $27$   &   $0$ & $523$ \\
\cline{1-6} \cline{8-13} \cline{15-20}\multicolumn{1}{|r @{\hspace{\dotlen}}}{$0$} &    &   \multicolumn{1}{r @{\hspace{\dotlen}}}{$0$} &    &   $-0$ & $214$   & &   $0$ & $122$   &   $0$ & $122$   &   $0$ & $624$   & &   $-0$ & $21$   &   $-0$ & $099$   &   $0$ & $398$ \\
\cline{1-6} \cline{8-13} \cline{15-20}$0$ & $24$   &   $0$ & $24$   &   \multicolumn{1}{r @{\hspace{\dotlen}}}{$-1$} &    & &   $0$ & $052$   &   $0$ & $399$   &   $0$ & $598$   & &   $-0$ & $224$   &   $-0$ & $224$   &   $0$ & $316$ \\
\cline{1-6} \cline{8-13} \cline{15-20}$0$ & $399$   &   $-0$ & $052$   &   $-0$ & $598$   & &   $-0$ & $152$   &   $0$ & $152$   &   $0$ & $472$   & &   $-0$ & $127$   &   $-0$ & $127$   &   $0$ & $269$ \\
\cline{1-6} \cline{8-13} \cline{15-20}$0$ & $152$   &   $0$ & $152$   &   $-0$ & $472$   & &   $0$ & $27$   &   $0$ & $27$   &   $0$ & $523$   & &   $-0$ & $158$   &   $0$ & $027$   &   $0$ & $263$ \\
\cline{1-6} \cline{8-13} \cline{15-20}$0$ & $21$   &   $-0$ & $099$   &   $-0$ & $398$   & &   \multicolumn{1}{r @{\hspace{\dotlen}}}{$0$} &    &   \multicolumn{1}{r @{\hspace{\dotlen}}}{$0$} &    &   $0$ & $388$   & &   $-0$ & $052$   &   $-0$ & $399$   &   $0$ & $598$ \\
\cline{1-6} \cline{8-13} \cline{15-20}$0$ & $158$   &   $0$ & $027$   &   $-0$ & $263$   & &   $0$ & $099$   &   $0$ & $21$   &   $0$ & $398$   & &   $-0$ & $099$   &   $-0$ & $21$   &   $0$ & $398$ \\
\end{longtable}

\end{document}